\documentclass[aps,prd,twocolumn,nofootinbib,superscriptaddress]{revtex4-2}
\usepackage{amsmath}
\usepackage{amssymb}
\usepackage{amsfonts}
\usepackage{balance}
\usepackage{booktabs, array}
\usepackage{multirow}
\usepackage{tikz}
\usepackage{graphicx} 
\usepackage{array} 
\usepackage[]{xcolor}
\usepackage{dcolumn}
\usepackage{ulem}
\usepackage{threeparttable}
\usepackage{comment}
\def\chain#1#2{\mathrel{\mathop{\null\longrightarrow}\limits^{#1}_{#2}}}

\begin{document}

\title{Fundamental quantum limits for detecting ultrahigh frequency
    gravitational waves}

\author{Xinyao Guo}
\affiliation{Department of Physics, Tsinghua University, Beijing 100084, China}
\author{Haixing Miao}
\affiliation{Frontier Science Center for Quantum Information, Department of Physics, Tsinghua University, Beijing 100084, China}
\author{Zhi-Wei Wang}
\affiliation{School of Physics, The University of Electronic Science and Technology of China, 88 Tian-run Road, Chengdu, China}
\author{Huan Yang}
\email{hyangdoa@tsinghua.edu.cn}
\affiliation{Department of Astronomy, Tsinghua University, Beijing 100084, China}
\author{Ye-Ling Zhou}
\affiliation{School of Fundamental Physics and Mathematical Sciences, Hangzhou Institute for Advanced
Study, UCAS, Hangzhou 310024, China}
\begin{abstract}
The ultrahigh-frequency (above 10 kHz) gravitational waves (GW) window provides a unique opportunity to detect primordial GWs, free from astrophysical foregrounds that dominate lower frequencies. A stochastic GW background in this range is generically predicted from cosmological phase transitions and topological defects associated with grand unification and other ultra-high energy theories. We establish a universal quantum limit framework for various detection schemes, setting a fundamental bound on GW detectability. Our analysis reveals that backgrounds in the kHz – MHz range are in principle observable, whereas higher-frequency signals lie below the quantum limit. These results offer theoretical guidance for future detector designs and open new avenues for probing early universe physics.
\end{abstract}
\maketitle

\section{Introduction}



Since the first detection of the binary black hole merger event GW150914 \cite{LIGOScientific:2016aoc}, the LIGO-Virgo-KAGRA  collaboration has achieved tremendous success in detecting more than 100 compact binary mergers in the audio band \cite{KAGRA:2021vkt}. The most recent Pulsar Timing Array measurements also show promising tentative evidence of a gravitational wave background in the ${\rm n}$Hz band \cite{NANOGrav:2023gor,EPTA:2023fyk,Reardon:2023gzh,Xu:2023wog}. In the next decade, spaceborne detectors such as LISA (Laser Interferometer Space Antenna), Taiji, and Tianqin \cite{Colpi:2024xhw,Hu:2017mde,luo2016tianqin} are likely to detect gravitational waves in the ${\rm m}$Hz range. Given these advances, it is now timely to investigate GW signals in the ultra-high frequency regime and explore the best observation strategies for such detections.

\begin{figure*}[t] 
\centering
\includegraphics[width=.8\textwidth]{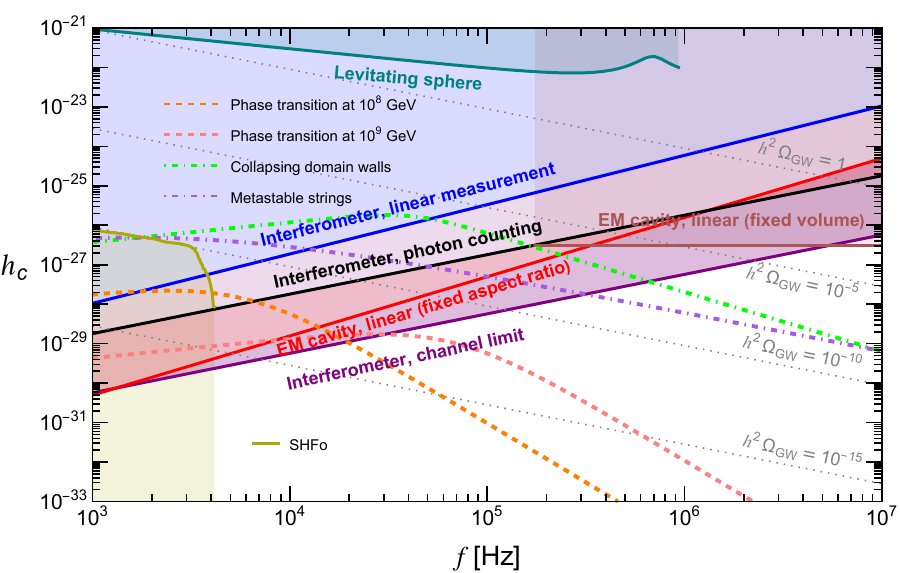}
\caption{Fundamental quantum limits for different experimental setups over $10^3$–$10^7$~Hz. \textcolor{black}{\textcolor{black}{Here, the integration time $T_{\rm int}$ is set to 10 years, {and the frequency bin width is set as the entire resonant bandwidth of individual detector}. {The EM cavity (fixed ratio) line assumes a cylindrical cavity with aspect ratio of 1:5, while benchmark values of other sensitivity lines are summarized in Tab.\,\ref{table:scaling}.}}
}  A representative binary neutron-star foreground (equation of state of SHFo \cite{Steiner:2012rk}, data extracted from \cite{L-shaped}) rapidly decays above $4$~kHz. The UHF band is therefore well-suited to probe GUT-motivated phase transitions (B1 at $10^8$~GeV, B2 at $10^9$~GeV, {with specific lineshape given in Eq.\,\ref{eq:GWsignal}}), collapsing domain walls (B3), or metastable strings (B4). }
\label{fig:limits}
\end{figure*}



The ultra-high frequency gravitational wave (GW) band ($\geq 10$ kHz) provides a ``golden window" to probe the early universe. GWs in this range are generically predicted in particle physics scenarios, arising from cosmological phase transitions, topological defects, and other primordial sources (see, e.g., reviews in \cite{Aggarwal:2020olq}). Unlike lower-frequency bands dominated by astrophysical foregrounds, this regime is largely free from contamination. In particular, the characteristic frequencies of dynamical processes involving astrophysical black holes and neutron stars are limited to $\lesssim 10$ kHz \cite{Shibata:2019ctb,Godzieba:2020tjn,Alsing:2017bbc}, meaning that gravitational wave backgrounds from such sources diminish significantly above this range. Consequently, this suppression removes a major barrier to detecting primordial GW signals, making the ultra-high frequency band especially promising for fundamental physics \cite{Pan:2023naq}.


Laser interferometers have reached quantum-limited measurement accuracy and have proven extremely successful in detecting gravitational waves in the audio band. For ultrahigh frequencies $\ge 10~{\rm kHz}$, in addition to the laser interferometer, there are also several proposals using microwave cavities and cavity-assisted levitating spheres. However, it is yet unclear what the optimal design and the best measurement accuracy would be. In order to guide future studies in this direction, i.e. what kind of theoretical models can be tested in this frequency band, we provide a unified framework to determine the ultimate measurement precision from a quantum-limited measurement perspective. With such a framework, we have considered various viable proposals and set their ultimate measurement precision, with all classical noises excluded. The main result is given in Fig.~\ref{fig:limits}.  Gravitational waves under such limit are likely {\it not} detectable based on current understanding. As an application of such detection limit, we analyze major Grand Unification Model(s) and present the parameter range that may be tested by detectors saturating the fundamental limit. 

{This paper is organized as follows. In Sec.\ref{sec:limits}, we propose a universal fundamental sensitivity limit applicable to a broad class of detectors. In Sec.\ref{sec:design}, we present a worked example illustrating how the fundamental limit can guide practical detector design. In Sec.\ref{sec:source}, we summarize the key theories that predict strong stochastic gravitational-wave signals at ultrahigh frequencies. Finally, we conclude in Sec.\ref{sec:conclusion}. Detailed derivations and proofs of the main results are provided in the appendices.}


\section{Fundamental limits}\label{sec:limits} 
Despite apparent differences in physical layouts, all proposed high-frequency detectors can be unified within a unified framework due to their shared mathematical description of ultimate sensitivity and their common core: the electromagnetic resonator.

Mathematically, detection sensitivity can be treated as a quantum estimation problem\,\cite{Stochastic}. A stochastic GW signal induces a random shift in the detector's quantum state, quantified by the normalized excess variance:
\begin{equation}
\sigma^2(f)=|G(f)|^2\,{S_{hh}(f)}\,,
\end{equation}
where the gain $G$ quantifies detector's response to the
GW strain $h$, {and $S_{hh}(f)$ is the single-sided power spectral density of strain.} The accuracy for estimating the excess variance, quantified by the signal-to-noise ratio (SNR), is determined by both the input quantum state and measurement scheme on the out-going quantum state. SNR limits for different schemes are summarized in Table\,\ref{tab:comparison}. {
In this table, the SNR limit for photon-counting metrology applies only to systems without internal squeezing, whereas the other SNR limits remain valid even in the presence of internal squeezing (See Appendix \,\ref{sec:loss} and Appendix \,\ref{sec:QFI} for detailed discussion).}

Physically, in all detectors, GWs couple linearly to the system's total stored energy in the conventional transverse-traceless (TT) gauge, with interaction Hamiltonian of:
\begin{equation}
\hat H_{\rm int} = {a}\, h\, \hat{\cal E}_{EM}\,,
\end{equation}
where $h$ is the strain, ${\cal E}_{EM}$ the EM energy, and $a$ the antenna response, an order-one factor for sky-averaged coupling. \textcolor{black}{For the linear, Gaussian, lossless case, strain sensitivity for detection is bounded by quantum energy fluctuations (energetic quantum limit \cite{EQL1,EQL2,EQL3,EQL4,Haixing2017}). {When the internal squeezing negligible}, we generalize this as a gain constraint applicable to general lossy, non-Gaussian cases:}
\begin{equation}
\label{eq:gain}
|G(f)|^2 = {a}^2\frac{\bar{S}_{\cal EE}^{\,\rm vac}(f)}{\hbar^2}\,.
\end{equation}
\textcolor{black}{With the spectral density of energy fluctuations $\bar{S}_{\cal EE}^{\,\rm vac}$ is given by:}
\begin{equation}\label{eq:lineshape}
\bar{S}^{\text {vac}}_{\cal E E}(f) = \bar{\cal E}\sum_k S_k^{\rm vac}= \bar{\cal E}\sum_k \frac{2\hbar (\omega_0 + \Delta_k ) \gamma_k}{(2\pi f-|\Delta_k|)^2 + \gamma_k^2}\,,
\end{equation}
which is a summation of Lorentzian spectra with {resonant frequency $\Delta_k$} linewidth $\gamma_k$, and total energy $\bar {\cal E}$. {Here, the summation run over all possible resonant frequencies of cavity modes. Detailed derivations of Eqs.\,\eqref{eq:gain} and \eqref{eq:lineshape} are summarized in Appendix \,\ref{sec:gain} and Appendix \,\ref{sec:energy}, respectively.}

\begin{table*}[t!]
\centering
\renewcommand{\arraystretch}{1.5} 
\setlength{\tabcolsep}{3pt} 
\small 
\begin{tabular}{|>{\centering\arraybackslash}m{1.4cm}|
>{\centering\arraybackslash}m{2.1cm}|
>{\centering\arraybackslash}m{1.6cm}|
>{\centering\arraybackslash}m{2.0cm}|
>{\centering\arraybackslash}m{1.9cm}|
>{\centering\arraybackslash}m{2.2cm}|
>{\centering\arraybackslash}m{1.3cm}|
>{\centering\arraybackslash}m{1.5cm}|
>{\centering\arraybackslash}m{1.8cm}|}
\hline
\(\raisebox{1.2ex}{\textbf{System}}\) & 
\(\raisebox{1.2ex}{\textbf{Measurement}}\) & 
\(\raisebox{1.2ex}{\textbf{SNR}}\) & 
\(\raisebox{1.2ex}{$\mathbf{h_{c,\text{min}}^{2}(f)}$}\) &  
{\textcolor{black}{\textbf{Measured Quantity}}} &{\textcolor{black}{\textbf{Estimator (for \(\sigma^2\))}}} &
{\textbf{Optimal Size $L$}} &
\(\raisebox{1.2ex}{\textbf{Energy $\bar{\mathcal{E}}$}}\) & \textbf{Bandwidth
$\gamma(f)$}\\
\hline
\multirow{3}{*}[10pt]{Laser}  
& \textcolor{black}{homodyne detection}& \(\raisebox{0.8ex}{\({ \sigma^2} {\varepsilon}^{-1} \sqrt{N}\)}\) 
& \(\raisebox{1.5ex}{\(\varepsilon S_n^{\rm vac}(f) N^{-\frac{1}{2}}\)}\) 
& \(\raisebox{1.5ex}{\(\Delta{\rm P}\propto \sigma\)}\) & \(\raisebox{1.5ex}{\(S_{\rm PP}({\hbar \omega_0 \rm P_{\rm LO})^{-1}}\)}\) & \multirow{3}{*}[-3pt]{$c / 4f$} & \multirow{3}{*}[-3pt]{${2{\rm P}_{\rm arm}L}/{c}$} &  \multirow{3}{*}[-3pt]{$2{\rm T}_{\rm src}f$} \\
\cline{2-4} \cline{5-6}
\(\raisebox{1.2ex}{Inter-}\)& photon counting & \(\raisebox{1.2ex}{\({\sigma } \sqrt{N{/2}}\)}\) 
& \(\raisebox{1.2ex}{\(2 S_n^{\rm vac}(f) N^{-1}\)}\) 
& \(\raisebox{1.5ex}{\(n_{\rm tot}= N \sigma^2/2\)}\) & \(\raisebox{1.2ex}{\(2n_{\rm tot}/N\)}\) &  &  & \\
\cline{2-4} \cline{5-6}
\(\raisebox{4ex}{ferometer}\)& \(\raisebox{1.2ex}{optimal}\)  & \(\raisebox{1.2ex}{\({\sigma }{\varepsilon}^{-\frac{1}{2}} \sqrt{N{/2}}\)}\) 
& \(\raisebox{1.2ex}{\(2 \varepsilon S_n^{\rm vac}(f) N^{-1}\)}\) 
& \(\raisebox{1.2ex}{unclear}\) & \(\raisebox{1.2ex}{unclear}\) &  & & \\
\hline
EM Cavity & \(\raisebox{1.5ex}{linear}\) & \(\raisebox{1.5ex}{\(\sigma^2 \sqrt{N}\)}\) 
& \(\raisebox{1.8ex}{\( S_n^{\rm vac}(f) (N)^{-\frac{1}{2}}\)}\) 
& \(\raisebox{1.5ex}{\({{\rm P}_{\rm tot } \propto \sigma}\)}\) & \(\raisebox{1.5ex}{\({{S}_{\rm PP}(f)({\hbar \omega^2 \bar{\cal E}})^{-1}}\)}\) & \(\raisebox{1.2ex}{$c / f$}\) & \(\raisebox{1.2ex}{${B_0^2}V/{2\mu_0}$}\) & \(\raisebox{1.2ex}{Constant $\gamma_0$}\) \\
\hline
Levitating Sphere & \(\raisebox{1.5ex}{linear}\) & \(\raisebox{1.5ex}{\({ \sigma^2} {\varepsilon}^{-1} \sqrt{N}\)}\) 
& \(\raisebox{1.8ex}{\( \varepsilon S_n^{\rm vac}(f) (N)^{-\frac{1}{2}}\)}\) 
& \(\raisebox{1.5ex}{\(\Delta{\rm P}\propto \sigma \)}\) & \(\raisebox{1.5ex}{\(S_{\rm PP}({\hbar \omega_0 \rm P_{\rm LO})^{-1}}\)}\) & \(\raisebox{1.2ex}{$\gg c / 2f$}\) & \(\raisebox{1.2ex}{${2{\rm P}_{\rm arm}L}/{c}$}\) & \(\raisebox{1.2ex}{Constant $\gamma_0$}\) \\
\hline
\end{tabular}
\caption{Detection capability of different proposals for weak stochastic signals, together with their estimators and key physical quantities. Here,
$S_n^{\rm vac}(f)=\frac{f \,\hbar^2}{{a}^2(f)\cdot \bar{S}_{\cal EE}^{\rm vac}(f)}$ is the vacuum shot noise level,
$\varepsilon=\varepsilon(f)$ is the system’s loss level ({depolarizing probability}), and $N=T_{\rm int}\,\Delta f$ {is the effective number of samples for estimating the magnitude of single-sided strain spectrum} .  Among these designs, photon counting and EM cavity are loss-insensitive, and other schemes are limited by quantum loss. 
For estimator part, $P$ and $P_{\rm LO}$ are the readout and local oscillator powers; $n_{\rm tot}$ is the total discrete signal photon number in photon counting, obeying Poisson statistics; $P_{\rm tot}$  is electromagnetic field power in the resonant cavity.  
For key physical quantities, ${\rm P}_{\rm arm}$ is the circulating power in the arm cavity, $B_0$ and $V$ are the static magnetic field and cavity volume, and $T_{\rm src}$ denotes the effective transmissivity of the signal-recycling cavity in the interferometer. Meanwhile, at specific frequency, only detector with optimal size saturates the fundamental limit. }
\label{tab:comparison}
\end{table*}

With the physical constraint on gain, {for any given integration time $T_{\rm int}$ and frequency bin width $\Delta f$, the SNR limit yields} the minimum detectable characteristic strain for unity SNR:
\begin{equation}
h^2_{\rm c, min}(f)\equiv f\cdot S_{hh}(f)|_{\rm SNR=1}\,.
\end{equation}
Minimum detectable signals for different designs are summarized in Tab.~\ref{tab:comparison}. 
{Meanwhile, in particle cosmology, a commonly used parameter to characterize the GW intensity is the GW energy-density spectrum 
$\Omega_{\mathrm{GW}}(f)\,,$
defined as the GW energy density per logarithmic frequency interval.}
It relates to the characteristic strain via: $\Omega_{\rm{GW}}(f)=2\pi^2/(3H_0^2)f^2 h_c^2(f)$, where $H_0 = 100 {\rm h} ~{\rm km/s/Mpc}$ is the Hubble constant. 

{For a realistic single detector, the maximum effective frequency bin width is physically constrained by the linewidth of resonant mode. Thus, the $h_{\rm c,min}$ with the choice of  $\Delta f= \gamma/ 2\pi$ represents the fundamental sensitivity limit of single detector. Numerical results of the limit are summarized in Fig.~\ref{fig:limits}. Stochastic signals below these thresholds are not likely to be detected by single detector.}


\textcolor{black}{The fundamental limits are closely connected to the state-of-the-art practical parameters. Scaling behavior of the sensitivity of each designs with respect to practical parameters are summarized in Tab.\,\ref{table:scaling}.}

\begin{table}[htbp]
\centering
\renewcommand{\arraystretch}{1.5} 
\setlength{\tabcolsep}{3pt}       
\begin{tabular}{|>{\centering\arraybackslash}m{2.0cm}
                |>{\centering\arraybackslash}m{2.3cm}
                |>{\centering\arraybackslash}m{3.7cm}|}
\hline
\(\raisebox{1.2ex}{\textbf{Scheme}}\) & \textbf{Relevant Parameters} & \(\raisebox{1.2ex}{\textbf{$\mathbf{h_{\rm c,min}^2}$ Scaling}}\) \\
\hline
Interferometer Linear & \(\raisebox{1.2ex}{${\rm P}_{\rm arm},\,\, {\varepsilon(f)},\,\,{\rm T}_{\rm src}$}\) & \(\raisebox{1.2ex}{$ \left(\frac{10 \mathrm{MW}}{\rm P_{\mathrm{arm}}}\right)\left(\frac{{\varepsilon(f)}}{0.001}\right) \left(\frac{\rm T_{\rm src}}{0.01}\right)^{-\frac{1}{2}}$}\) \\
\hline
Interferometer Counting & \(\raisebox{1.2ex}{${\rm P}_{\rm arm}$}\) & \(\raisebox{1.2ex}{$ \left(\frac{10 \mathrm{MW}}{\rm P_{\mathrm{arm}}}\right)$}\) \\
\hline
Interferometer Channel limit & \(\raisebox{1.2ex}{${\rm P}_{\rm arm},\,\, {{\varepsilon(f)}}$}\) & \(\raisebox{1.2ex}{$\left(\frac{10 \mathrm{MW}}{\rm P_{\mathrm{arm}}}\right)\times \left(\frac{{\varepsilon(f)}}{0.001}\right) $}\) \\
\hline
EM cavity & {${\rm B}_{0},\,\, \gamma_0$}& {$ \left(\frac{5 \mathrm{T}}{\rm B_{0}}\right)^2\left(\frac{\gamma_0}{1 \rm{kHz}}\right)^{\frac{1}{2}} {\left(\frac{[300 \,\rm m]^3}{V}\right)}$} \\
\hline
Levitating Sphere& \(\raisebox{1.2ex}{$m_{\rm lev},\,\, \gamma_0$}\) & \(\raisebox{1.2ex}{$ \left(\frac{37.5 {\rm \mu g}}{m_{\rm lev}}\right) \times\left(\frac{\gamma_0}{1 \rm{kHz}} \right)^{\frac{1}{2}}$}\) \\

\hline
\end{tabular}
\caption{\textcolor{black}{Parameter dependence and scaling of the fundamental limit for different designs. 
Here, {$V=(300 \,\rm m)^3$ is the benchmark value for the fixed volume EM cavity,} $m_{\rm lev}$ is the mass of the levitated object inside the cavity, and meaning of other parameters is the same as in Table~\ref{tab:comparison}. 
}
}
\label{table:scaling}
\end{table}
\textcolor{black}{Notably, these limits are also compatible with cross-correlation analysis. $h_{\rm c,min}$ of individual detector is equivalent to that of two co-located identical detectors up to a factor of $\sqrt{2}$\,\cite{cross_lin, photoncounting}, and differs from cross-correlation between two distant detectors only by the overlap reduction factor\,\cite{cross_lin}. }
\section{FQL-based conceptual design}\label{sec:design}
The fundamental quantum limit serves as a direct guideline for systematic, conceptual-level detector design. To illustrate this process, we present a worked example focused on detecting the collapsing domain-wall signal at approximately $75\,\mathrm{kHz}$.

\begin{figure*}[t]
\centering
\includegraphics[width=.41\textwidth]{ 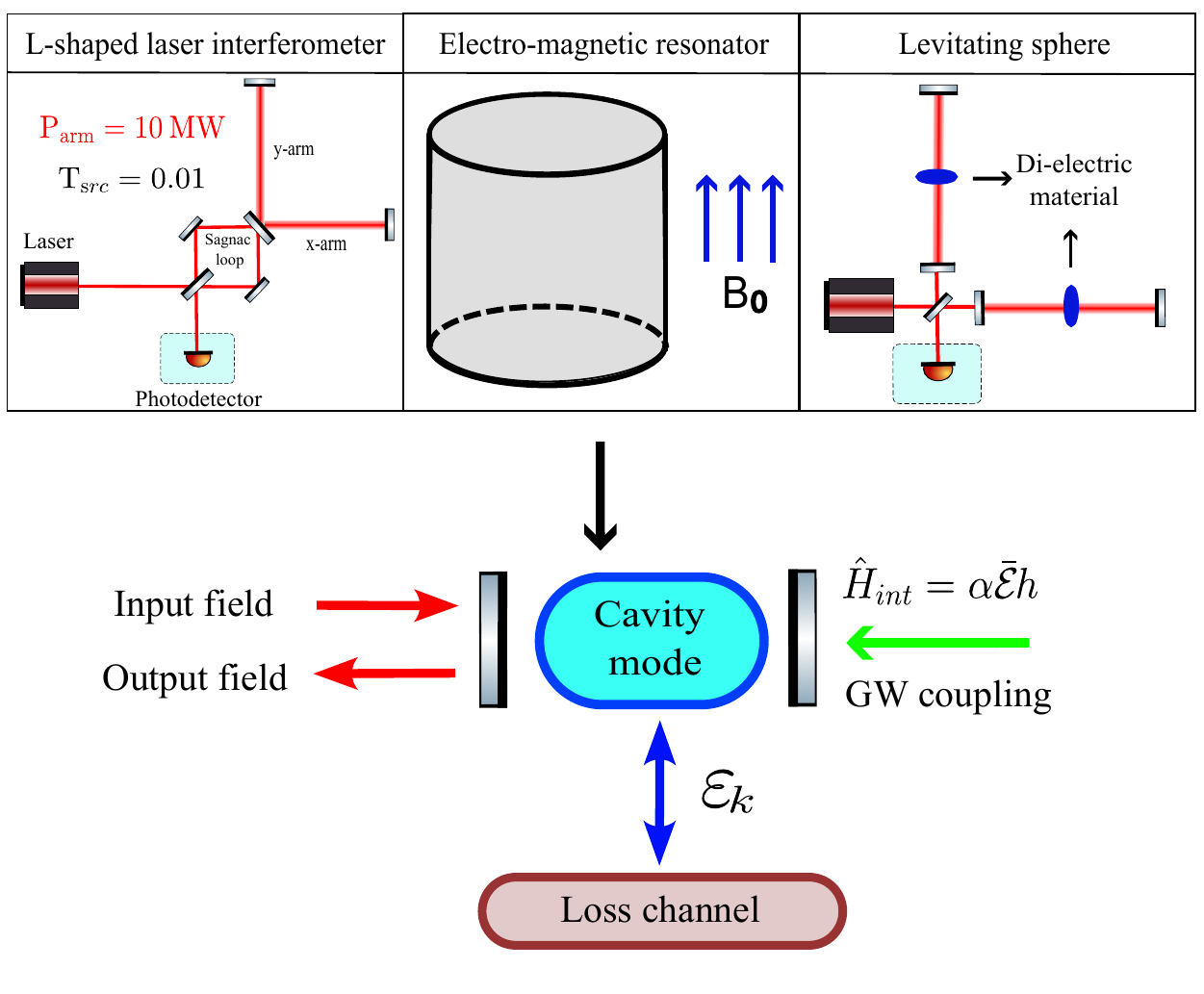}\,
\includegraphics[width=.47\textwidth]{ 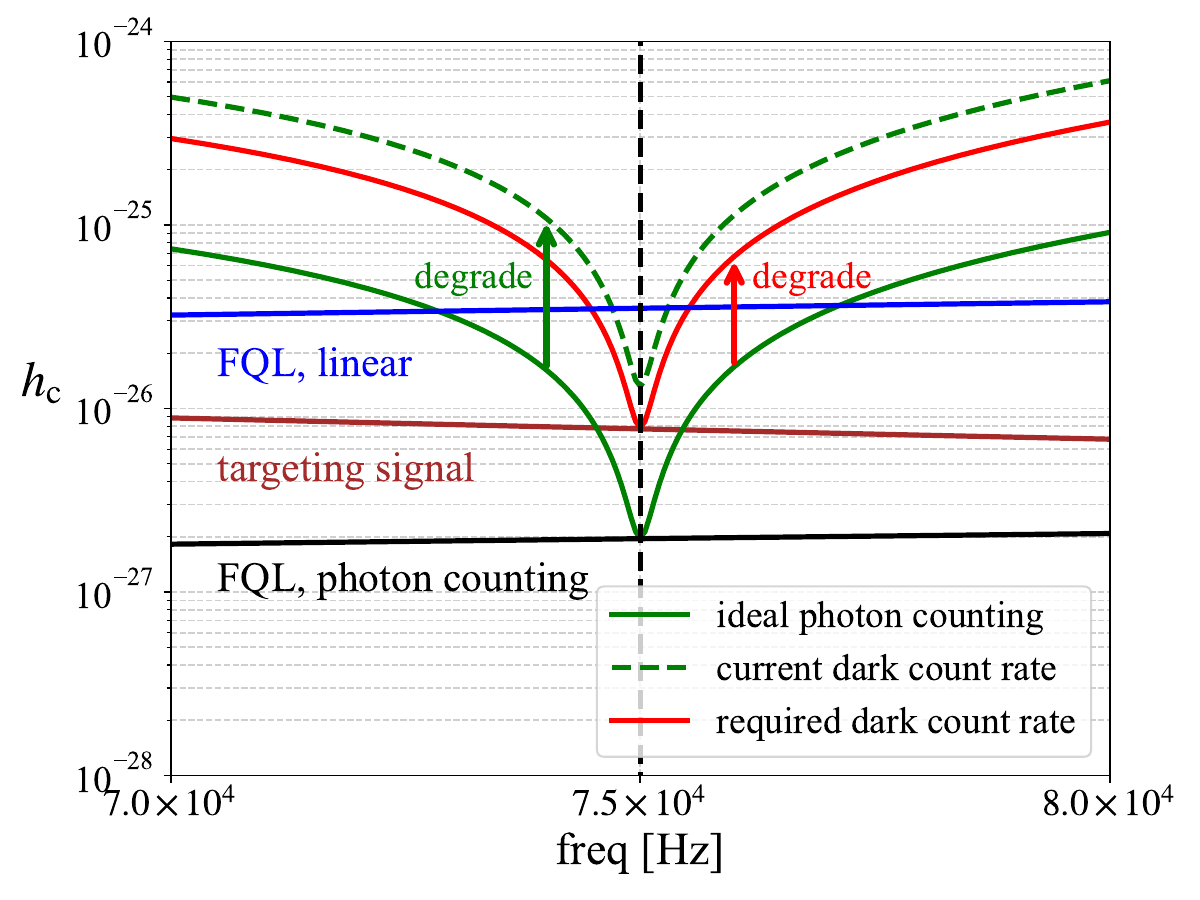}\label{fig:corner1}
\caption{\textcolor{black}{Left: schematic diagram of concept and layout of detection proposals in Fig\,\ref{fig:limits}. Right:  Fundamental limits and single-detector sensitivity under different classical noise levels, corresponding to the worked example of fundamental-limit-based detector design.}}
\label{fig:example}
\end{figure*}

The first step is to select the most suitable physical platform with the proper metrology using the fundamental limit. From Fig.\,\ref{fig:limits}, both laser interferometers and electromagnetic (EM) cavities emerge as promising candidates. However, Table~\ref{tab:comparison} shows that detecting a $75\,\mathrm{kHz}$ signal would require a kilometer-scale detector, which is impractically large for 3-dimensional EM cavity. In addition, photon-counting measurements outperform linear homodyne detection. Taken together, these considerations make the laser interferometer with photon-counting readout the most suitable platform.

The next step is to identify the key parameters—total energy, bandwidth, loss, and antenna response—by analyzing the scaling behavior of the fundamental limits summarized in Table~\ref{table:scaling}. For photon-counting, the sensitivity is independent of bandwidth and loss, and is determined solely by the arm cavity power (i.e., the total energy). Thus, optimizing these parameters does not produce a unique working point. We adopt a nominal value of $10\,\mathrm{MW}$ for this parameter. Meanwhile, according to Ref.~\cite{L-shaped}, interferometer with L-shaped optical cavity provides the optimal antenna response. As illustrated in Fig.~\ref{fig:example}, the L-shaped interferometer design reaches the fundamental limit at $75\,\mathrm{kHz}$.

Finally, the impact of classical noise, which would further degrade sensitivity must be considered. In this example, the dominant contribution arises from the residual photon flux $\bar{\dot n}_{0}$. Figure~\ref{fig:example} illustrates how the residual photon flux degrades the sensitivity. Reaching the targeting sensitivity requires $\bar{\dot n}_{0} \leq 8.0 \times 10^{-7}\,\mathrm{s}^{-1}$, whereas the current state-of-the-art level is about $7.0 \times 10^{-6}\,\mathrm{s}^{-1}$, as reported in Ref.~\cite{axion}.  This comparison highlights the gap between present performance and the fundamental limit, underscoring the need for further suppression of classical noise.


\section{Implications}\label{sec:source}
Based on the quantum-limited sensitivity on high-frequency GW detection, we discuss the implications on how well one can test some of the well-motivated predictions related to Grand Unified Theories (GUTs) and topological defects (domain walls and cosmic strings).

\begin{figure}[t!]
\centering
\includegraphics[width=.4\textwidth]{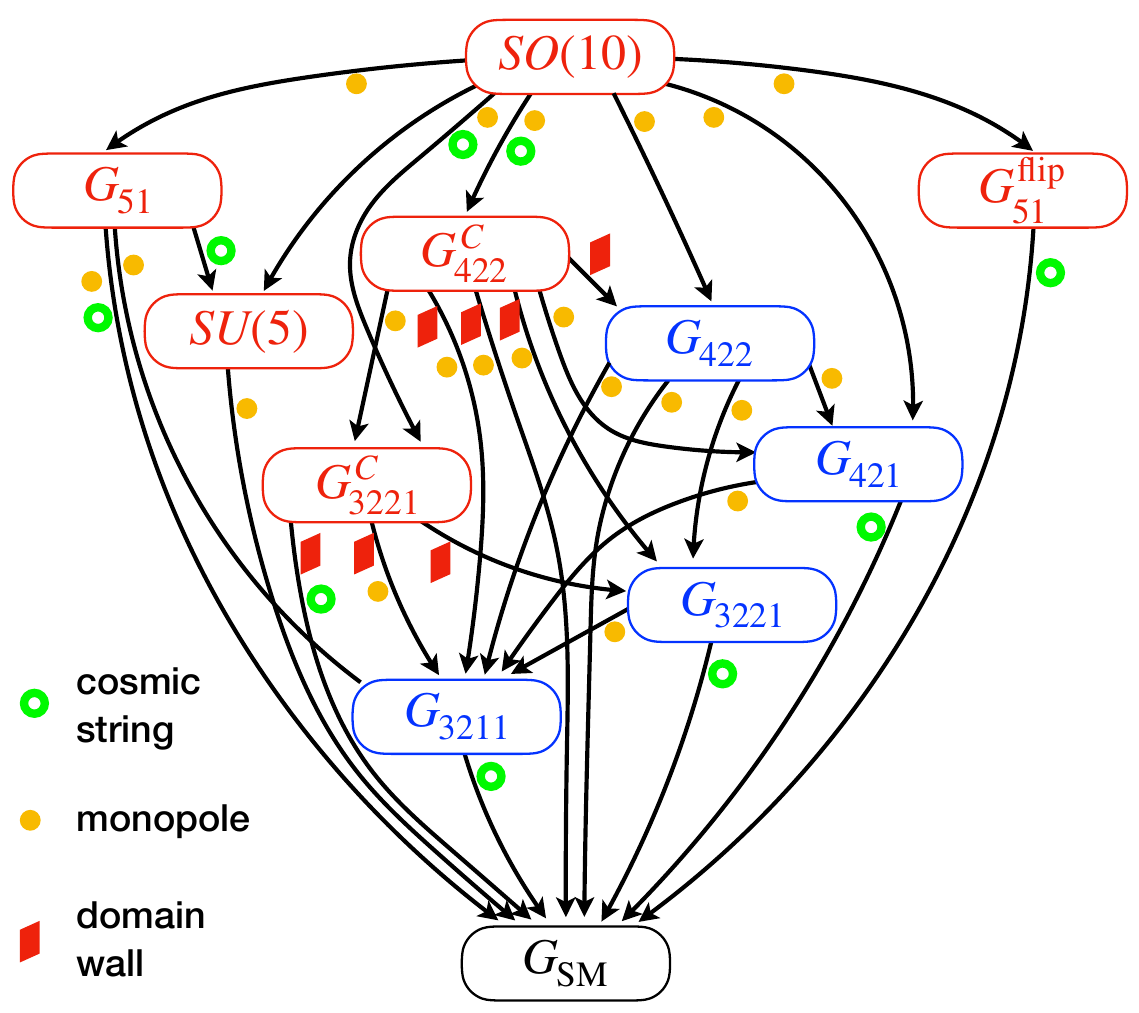}
\caption{Non-SUSY $SO(10)$ GUT breaking chains as motivations of ultra-high frequency GWs via phase transition. Conventions, e.g., $G_{422} = SU(4)_c\times SU(2)_L\times SU(2)_R$, are understood \cite{King:2021gmj}. $G_{\rm SM}\equiv SU(3)\times SU(2)_L \times U(1)_Y$ denotes the gauge symmetry of the Standard Model. Topological defects (cosmic strings, monopoles and domain walls) associated with the symmetry breaking \cite{Jeannerot:2003qv} are marked. Gauge symmetries which might provide phase transition at scales lower than $10^{10}$~GeV and consistent with cosmological observations are highlighted in blue. \label{fig:GUT}}
\end{figure}

\textbf{GUT-motivated phase transition:} GUTs, which aim to unify the three fundamental particle forces, can naturally predict cosmological phase transitions at ultra-high-energy scales. Although the energy scale of GUT breaking ($\gtrsim 10^{16}$~GeV) is prohibitively high, intermediate symmetry breaking can naturally occur at lower scales but sufficiently higher than the electroweak scale. {\color{black} Below, we will focus on non-SUSY $SO(10)$ GUTs \cite{Fritzsch:1974nn} as our benchmark because this framework is among the best classified, enabling an exhaustive audit of viable breaking chains and a direct mapping to our quantum-limited detectability criterion. It is worth mentioning that SUSY GUTs \cite{Aulakh:1982sw} or GUTs with larger gauge symmetries, e.g., $E_6$ \cite{Gursey:1975ki}, provide new recipes of both symmetries and extra DOFs. They modify details of dynamics without altering the qualitative result for the study of phase transitions and thus will not be considered here. The $SO(10)$ breaking chains via certain intermediate symmetries to the SM is summarized in Fig.~\ref{fig:GUT}. The relevant intermediate scales are $10^6$–$10^{10}$~GeV; the blue chains in Fig.~\ref{fig:GUT} fall in this range and can source ultra-high-frequency (UHF) GWs.} 
Two representative benchmarks of GW spectra at $10^{8}$~GeV (B1) and $10^{9}$~GeV (B2) are shown in the right panel of Fig.~\ref{fig:limits}, which can result in typical symmetry breaking $G_{422} \to G_{3221}$ or $G_{3221} \to G_{\rm SM}$ shown in Fig.~\ref{fig:GUT}. Earlier results of GWs induced by Pati-Salam symmetry breaking without embedding in a full $SO(10)$ framework can be found in \cite{Croon:2018kqn,Huang:2020bbe}.

The intermediate symmetry breaking can be triggered by a Higgs-like mechanism, involving a scalar field that acquires a vacuum expectation value (VEV). 
\textcolor{black}{It entails a high-temperature cosmological phase transition at $T_*$ in the early universe. Its dynamics follow from an effective potential with zero- and finite-temperature loop corrections from gauge bosons, scalars, and fermions participating in the transition. Once the model is specified, the potential is fixed. Connections between phase transition parameters and GUT particle models will be discussed in \cite{new:ref}.}
During a first-order phase transition, the energy stored in the false vacuum is drastically released into the bulk through bubble nucleation, generating gravitational radiation. The latter contributes as a stochastic GW background observed today. For phase transitions that involve gauge-symmetry breaking, the sound waves in the bulk plasma are considered as the primary source of GWs. The acoustic GW power spectrum today is \cite{Hindmarsh:2017gnf,Weir:2017wfa}:
\begin{align}
 {\rm{h}}^2\Omega_\text{GW}(f)&= {\rm{h}}^2\Omega_\text{p}{\cdot}
 \Big(\frac{f}{f_\text{p}}\Big)^{\!3} \left[ \frac{4}{7}+\frac{3}{7}\Big( \frac{f}{f_\text{p}} \Big)^{\!2}\right]^{-\frac{7}{2}}, 
 \label{eq:GWsignal}
\end{align}
with the peak frequency and peak amplitude given by
\begin{align}
f_\text{p}&\simeq 89\,\text{kHz} \, v_w^{-1}\Big(\frac{g_*}{100} \Big)^{\!\frac{1}{6}}\Big( \frac{T_*}{10^9\, \text{GeV}}\Big) \Big(\frac{\tilde \beta}{10^3} \Big),
\nonumber\\
{\rm{h}}^2\Omega_\text{p} &\simeq 1.19\cdot 10^{-9} v_w \Big(\frac{10^3}{\tilde \beta}\Big)\Big( \frac{\kappa_{\rm{sw}}(\alpha)\, \alpha}{1+\alpha} \Big)^{\!2}\Big(\frac{100}{g_*}\Big)^{\!\frac{1}{3}}\,,\label{eq:peak-amp}
\end{align}
where {$\Omega_{\rm p}$ is a dimensionless scalar, and} $\rm{h}$ is the reduced Hubble parameter and it has nothing do to with the strain. Here, $T_*$ denotes the temperature at the phase transition, $v_w$ represents the bubble wall velocity, and $g_*$ characterizes the number of degrees of freedom (DOF) participating in the phase transition.
The peak amplitude is determined primarily by two parameters: $\alpha$, which quantifies the energy released into the plasma normalized by the radiation energy, and $\tilde\beta$, which represents the inverse duration of the phase transition. Furthermore, $\kappa_{\rm{sw}}(\alpha)$ is an efficiency factor dependent on $\alpha$, where an additional suppression factor due to the length of the period of sound waves is encoded\cite{Espinosa:2010hh,Guo:2020grp}. {Worth to point out, the frequency dependency of this GUT-motivated signal does not exhibit a common Lorentzian shape.} 


\textcolor{black}{A phase transition at $T_*\approx 10^7\,\mathrm{GeV}$ maps to the kHz GW band, while $T_*\approx 10^{10}\,\mathrm{GeV}$ maps to the MHz band. B1 becomes testable once sensitivities reach the quantum limit. For B2, the peak and UV lie below the quantum limit, while the IR tail remains within reach. Detecting only the IR tail would not fix the spectral shape; attribution to a phase transition would remain ambiguous. For intermediate breaking above  $10^{10}$~GeV, all bands fall below the quantum limit, precluding direct detection.}

\textbf{Collapsing domain walls:} Domain walls provide another mechanism for ultra-high-frequency gravitational waves. 
The domain wall is a topological defect that is inevitably generated after spontaneous breaking of discrete symmetries. The latter are frequently applied in beyond the Standard Model new physics models at ultra-high energy scales. 
Domain walls, once they are generated, appear as two-dimensional massive objects evolving in the universe. 
In order to avoid a domain-wall-dominated Universe, a bias term, originating from an asymmetry, is often introduced to break the degeneracy between vacuum states. They cause the walls to collapse during which period GWs are generated. The spectrum of GWs from collapsing domain walls follows broken power laws of $f^3$ on the IR band and roughly $f^{-1}$ on the UV band, and the peak frequency is determined by the annihilation temperature \cite{Saikawa:2017hiv}. Ultra-high-frequency GWs in the kHz-MHz band arise from discrete symmetry breaking at scales ranging from $10^9\,\rm{GeV}$ to the GUT scale, provided suitable bias terms are included \cite{Gelmini:2020bqg}. A benchmark of domain walls (B3) with tension (that is, surface energy) $\sigma \sim (10^{14}~{\rm GeV})^3$ is presented in the green curve in Fig.~\ref{fig:limits}. 
Collapsing domain walls are not predicted in GUTs. 
Some New Physics frameworks that naturally achieve it include: residual $Z_N$ symmetry $\sim 10^{12}$~GeV arising in axion models to address strong CP problem~\cite{Sikivie:1982qv} and discrete symmetries (either Abelian \cite{Wu:2022tpe} or non-Abelian \cite{Fu:2024jhu}) in solving the problem of quark and lepton flavor mixing~\cite{Xing:2020ijf,King:2017guk}. 

\textbf{Metastable strings:}
Cosmic strings are generated after spontaneous breaking of $U(1)_{B-L}$. The latter is naturally predicted in the Pati-Salam or relevant GUT frameworks \cite{Mohapatra:1980qe}, explaining the tiny mass of neutrino with Majorana nature in the seesaw framework \cite{Minkowski:1977sc,Yanagida:1980xy}. Cosmic strings form a network in the early Universe, intersecting to form loops that oscillate and emit GWs as they shrink \cite{Vilenkin:1984ib}. 
Simulations suggest that the GW spectrum from stable cosmic strings spans nHz to GHz, with a high-frequency plateau proportional to $\sqrt{\mu}$, where
$\mu$ is the string's energy density per unit length
\cite{Blanco-Pillado:2017oxo}. 
If the $B-L$ breaking scale is not hierarchically far below the GUT scale, the strings can decay to monopole-antimonopole pairs, forming a metastable string network \cite{Buchmuller:2019gfy}. GWs at lower frequencies, referring to those released from loops at later time, are suppressed because of the decay of strings. The decay width per string unit length is determined by the ratio $\kappa = m^2/ \mu$, where $m$ is the monopole mass of GUT naturally around the GUT scale. A smaller $\kappa$ means a faster decaying string network, corresponding to a higher frequency cutoff at the IR band. In Fig.~\ref{fig:limits}, we show a benchmark at $\kappa = 25$ (B4). The GW spectrum below $10^4$~Hz is suppressed and follows the power law $f^2$ until it reaches the plateau round $10^4$-$10^5$~Hz.

\section{Conclusion}\label{sec:conclusion}

In the ultra-high frequency band the universe is free of astrophysical stochastic GW foreground, which is certainly beneficial for probing primordial GWs. While the discussion of best detector configurations in this band is still in the early stage, we propose an unified framework to determine the quantum-limited sensitivity on strain, including proposals with varying coupling Hamiltonian, injected quantum states and readout scheme. The results obtained in Fig.~\ref{fig:limits} may be considered as the most optimistic bound that can be achieved, as classical noise sources are not included. With these bounds, we find that testing GUT theories with \textcolor{black}{intermediate-scale} phase transition temperature below $10^{10}\,\rm{GeV}$ is, in principle, allowed by future detectors in this band. Some of the domain-wall models with very large tension may also generate detectable GWs. 
 Crucially, our results provide a concrete theoretical foundation for future explorations of high-frequency GW sources, guiding both experimental strategies and theoretical investigations into early universe physics.

While preparing the manuscript, we find a related discussion on the sensitivity limit of high frequency detectors in \cite{dagnolo2024classical}, restricted to linear measurement in the ideal lossless case. 

\section{Acknowledgements}

X. G. and H. M. would like to acknowledge the support from the National Key R\&D Program of China under the project 'Gravitational Wave Detection' (Grant No.: 2023YFC2205800) and the support from the Frontier Science Center for Quantum Information. ZWW is supported in part by the National Natural Science Foundation of China (Grant No.~12475105). H.~Y. is supported by the Natural Science Foundation of China (Grant 12573048).  YLZ was partially supported by National Natural Science Foundation of China (NSFC) under Grant Nos. 12535007, 12547104, and Zhejiang Provincial Natural Science Foundation of China under Grant No. LDQ24A050002. ZWW and YLZ give special thanks to Z.Z. Xing for hospitality in IHEP where the initial collaboration was facilitated.

\bibliographystyle{unsrt}
\bibliographystyle{unsrt}
\bibliography{references}

@article{LIGOScientific:2016aoc,
    author = "Abbott, B. P. and others",
    collaboration = "LIGO Scientific, Virgo",
    title = "{Observation of Gravitational Waves from a Binary Black Hole Merger}",
    eprint = "1602.03837",
    archivePrefix = "arXiv",
    primaryClass = "gr-qc",
    reportNumber = "LIGO-P150914",
    doi = "10.1103/PhysRevLett.116.061102",
    journal = "Phys. Rev. Lett.",
    volume = "116",
    number = "6",
    pages = "061102",
    year = "2016"
}

@article{KAGRA:2021vkt,
    author = "Abbott, R. and others",
    collaboration = "KAGRA, VIRGO, LIGO Scientific",
    title = "{GWTC-3: Compact Binary Coalescences Observed by LIGO and Virgo during the Second Part of the Third Observing Run}",
    eprint = "2111.03606",
    archivePrefix = "arXiv",
    primaryClass = "gr-qc",
    reportNumber = "LIGO-P2000318",
    doi = "10.1103/PhysRevX.13.041039",
    journal = "Phys. Rev. X",
    volume = "13",
    number = "4",
    pages = "041039",
    year = "2023"
}

@article{NANOGrav:2023gor,
    author = "Agazie, Gabriella and others",
    collaboration = "NANOGrav",
    title = "{The NANOGrav 15 yr Data Set: Evidence for a Gravitational-wave Background}",
    eprint = "2306.16213",
    archivePrefix = "arXiv",
    primaryClass = "astro-ph.HE",
    doi = "10.3847/2041-8213/acdac6",
    journal = "Astrophys. J. Lett.",
    volume = "951",
    number = "1",
    pages = "L8",
    year = "2023"
}

@article{EPTA:2023fyk,
    author = "Antoniadis, J. and others",
    collaboration = "EPTA, InPTA:",
    title = "{The second data release from the European Pulsar Timing Array - III. Search for gravitational wave signals}",
    eprint = "2306.16214",
    archivePrefix = "arXiv",
    primaryClass = "astro-ph.HE",
    doi = "10.1051/0004-6361/202346844",
    journal = "Astron. Astrophys.",
    volume = "678",
    pages = "A50",
    year = "2023"
}

@article{Reardon:2023gzh,
    author = "Reardon, Daniel J. and others",
    title = "{Search for an Isotropic Gravitational-wave Background with the Parkes Pulsar Timing Array}",
    eprint = "2306.16215",
    archivePrefix = "arXiv",
    primaryClass = "astro-ph.HE",
    doi = "10.3847/2041-8213/acdd02",
    journal = "Astrophys. J. Lett.",
    volume = "951",
    number = "1",
    pages = "L6",
    year = "2023"
}

@article{Xu:2023wog,
    author = "Xu, Heng and others",
    title = "{Searching for the Nano-Hertz Stochastic Gravitational Wave Background with the Chinese Pulsar Timing Array Data Release I}",
    eprint = "2306.16216",
    archivePrefix = "arXiv",
    primaryClass = "astro-ph.HE",
    doi = "10.1088/1674-4527/acdfa5",
    journal = "Res. Astron. Astrophys.",
    volume = "23",
    number = "7",
    pages = "075024",
    year = "2023"
}

@article{Colpi:2024xhw,
    author = "Colpi, Monica and others",
    title = "{LISA Definition Study Report}",
    eprint = "2402.07571",
    archivePrefix = "arXiv",
    primaryClass = "astro-ph.CO",
    month = "2",
    year = "2024"
}

@article{luo2016tianqin,
  title={TianQin: a space-borne gravitational wave detector},
  author={Luo, Jun and Chen, Li-Sheng and Duan, Hui-Zong and Gong, Yun-Gui and Hu, Shoucun and Ji, Jianghui and Liu, Qi and Mei, Jianwei and Milyukov, Vadim and Sazhin, Mikhail and others},
  journal={Classical and Quantum Gravity},
  volume={33},
  number={3},
  pages={035010},
  year={2016},
  publisher={IOP Publishing}
}

@article{Hu:2017mde,
    author = "Hu, Wen-Rui and Wu, Yue-Liang",
    title = "{The Taiji Program in Space for gravitational wave physics and the nature of gravity}",
    doi = "10.1093/nsr/nwx116",
    journal = "Natl. Sci. Rev.",
    volume = "4",
    number = "5",
    pages = "685--686",
    year = "2017"
}

@article{Shibata:2019ctb,
    author = "Shibata, Masaru and Zhou, Enping and Kiuchi, Kenta and Fujibayashi, Sho",
    title = "{Constraint on the maximum mass of neutron stars using GW170817 event}",
    eprint = "1905.03656",
    archivePrefix = "arXiv",
    primaryClass = "astro-ph.HE",
    doi = "10.1103/PhysRevD.100.023015",
    journal = "Phys. Rev. D",
    volume = "100",
    number = "2",
    pages = "023015",
    year = "2019"
}

@article{Pan:2023naq,
    author = "Pan, Zhen and Yang, Huan",
    title = "{Improving the detection sensitivity to primordial stochastic gravitational waves with reduced astrophysical foregrounds}",
    eprint = "2301.04529",
    archivePrefix = "arXiv",
    primaryClass = "gr-qc",
    doi = "10.1103/PhysRevD.107.123036",
    journal = "Phys. Rev. D",
    volume = "107",
    number = "12",
    pages = "123036",
    year = "2023"
}

@article{Aggarwal:2020olq,
    author = "Aggarwal, Nancy and others",
    title = "{Challenges and opportunities of gravitational-wave searches at MHz to GHz frequencies}",
    eprint = "2011.12414",
    archivePrefix = "arXiv",
    primaryClass = "gr-qc",
    reportNumber = "CERN-TH-2020-185, HIP-2020-28/TH, DESY 20-195, CERN-TH-2020-185, HIP-2020-28/TH, DESY 20-195",
    doi = "10.1007/s41114-021-00032-5",
    journal = "Living Rev. Rel.",
    volume = "24",
    number = "1",
    pages = "4",
    year = "2021"
}

@article{Godzieba:2020tjn,
    author = "Godzieba, Daniel A. and Radice, David and Bernuzzi, Sebastiano",
    title = "{On the maximum mass of neutron stars and GW190814}",
    eprint = "2007.10999",
    archivePrefix = "arXiv",
    primaryClass = "astro-ph.HE",
    doi = "10.3847/1538-4357/abd4dd",
    journal = "Astrophys. J.",
    volume = "908",
    number = "2",
    pages = "122",
    year = "2021"
}

@article{Alsing:2017bbc,
    author = "Alsing, Justin and Silva, Hector O. and Berti, Emanuele",
    title = "{Evidence for a maximum mass cut-off in the neutron star mass distribution and constraints on the equation of state}",
    eprint = "1709.07889",
    archivePrefix = "arXiv",
    primaryClass = "astro-ph.HE",
    doi = "10.1093/mnras/sty1065",
    journal = "Mon. Not. Roy. Astron. Soc.",
    volume = "478",
    number = "1",
    pages = "1377--1391",
    year = "2018"
}

@article{Steiner:2012rk,
    author = "Steiner, Andrew W. and Hempel, Matthias and Fischer, Tobias",
    title = "{Core-collapse supernova equations of state based on neutron star observations}",
    eprint = "1207.2184",
    archivePrefix = "arXiv",
    primaryClass = "astro-ph.SR",
    reportNumber = "INT-PUB-12-033",
    doi = "10.1088/0004-637X/774/1/17",
    journal = "Astrophys. J.",
    volume = "774",
    pages = "17",
    year = "2013"
}

@misc{Stochastic,
    title={Stochastic waveform estimation at the fundamental quantum limit},
    author={James W. Gardner and Tuvia Gefen and Simon A. Haine and Joseph J. Hope and John Preskill and Yanbei Chen and Lee McCuller},
    year={2024},
    eprint={2404.13867},
    archivePrefix={arXiv},
    primaryClass={quant-ph}
}

@article{new:ref,
    author = "Xinyao Guo and Haixing Miao and Zhi-Wei Wang and Huan Yang and Ye-Ling Zhou",
    title = "{in progress}",}

@article{Hindmarsh:2017gnf,
    author = "Hindmarsh, Mark and Huber, Stephan J. and Rummukainen, Kari and Weir, David J.",
    title = "{Shape of the acoustic gravitational wave power spectrum from a first order phase transition}",
    eprint = "1704.05871",
    archivePrefix = "arXiv",
    primaryClass = "astro-ph.CO",
    reportNumber = "HIP-2017-02-TH, HIP-2017-02/TH",
    doi = "10.1103/PhysRevD.96.103520",
    journal = "Phys. Rev. D",
    volume = "96",
    number = "10",
    pages = "103520",
    year = "2017",
    note = "[Erratum: Phys.Rev.D 101, 089902 (2020)]"
}

@article{Guo:2020grp,
    author = "Guo, Huai-Ke and Sinha, Kuver and Vagie, Daniel and White, Graham",
    title = "{Phase Transitions in an Expanding Universe: Stochastic Gravitational Waves in Standard and Non-Standard Histories}",
    eprint = "2007.08537",
    archivePrefix = "arXiv",
    primaryClass = "hep-ph",
    doi = "10.1088/1475-7516/2021/01/001",
    journal = "JCAP",
    volume = "01",
    pages = "001",
    year = "2021"
}

@article{Georgi:1974sy,
    author = "Georgi, H. and Glashow, S. L.",
    title = "{Unity of All Elementary Particle Forces}",
    doi = "10.1103/PhysRevLett.32.438",
    journal = "Phys. Rev. Lett.",
    volume = "32",
    pages = "438--441",
    year = "1974"
}

@article{Fritzsch:1974nn,
    author = "Fritzsch, Harald and Minkowski, Peter",
    title = "{Unified Interactions of Leptons and Hadrons}",
    reportNumber = "CALT-68-467",
    doi = "10.1016/0003-4916(75)90211-0",
    journal = "Annals Phys.",
    volume = "93",
    pages = "193--266",
    year = "1975"
}

@article{King:2021gmj,
    author = "King, Stephen F. and Pascoli, Silvia and Turner, Jessica and Zhou, Ye-Ling",
    title = "{Confronting SO(10) GUTs with proton decay and gravitational waves}",
    eprint = "2106.15634",
    archivePrefix = "arXiv",
    primaryClass = "hep-ph",
    reportNumber = "IPPP/20/120",
    doi = "10.1007/JHEP10(2021)225",
    journal = "JHEP",
    volume = "10",
    pages = "225",
    year = "2021"
}

@article{Jeannerot:2003qv,
    author = "Jeannerot, Rachel and Rocher, Jonathan and Sakellariadou, Mairi",
    title = "{How generic is cosmic string formation in SUSY GUTs}",
    eprint = "hep-ph/0308134",
    archivePrefix = "arXiv",
    doi = "10.1103/PhysRevD.68.103514",
    journal = "Phys. Rev. D",
    volume = "68",
    pages = "103514",
    year = "2003"
}

@article{Saikawa:2017hiv,
    author = "Saikawa, Ken'ichi",
    title = "{A review of gravitational waves from cosmic domain walls}",
    eprint = "1703.02576",
    archivePrefix = "arXiv",
    primaryClass = "hep-ph",
    reportNumber = "DESY-17-036",
    doi = "10.3390/universe3020040",
    journal = "Universe",
    volume = "3",
    number = "2",
    pages = "40",
    year = "2017"
}

@article{Gelmini:2020bqg,
    author = "Gelmini, Graciela B. and Pascoli, Silvia and Vitagliano, Edoardo and Zhou, Ye-Ling",
    title = "{Gravitational wave signatures from discrete flavor symmetries}",
    eprint = "2009.01903",
    archivePrefix = "arXiv",
    primaryClass = "hep-ph",
    doi = "10.1088/1475-7516/2021/02/032",
    journal = "JCAP",
    volume = "02",
    pages = "032",
    year = "2021"
}

@article{Sikivie:1982qv,
    author = "Sikivie, P.",
    title = "{Of Axions, Domain Walls and the Early Universe}",
    reportNumber = "UFTP-82-3",
    doi = "10.1103/PhysRevLett.48.1156",
    journal = "Phys. Rev. Lett.",
    volume = "48",
    pages = "1156--1159",
    year = "1982"
}

@article{King:2017guk,
    author = "King, S. F.",
    title = "{Unified Models of Neutrinos, Flavour and CP Violation}",
    eprint = "1701.04413",
    archivePrefix = "arXiv",
    primaryClass = "hep-ph",
    doi = "10.1016/j.ppnp.2017.01.003",
    journal = "Prog. Part. Nucl. Phys.",
    volume = "94",
    pages = "217--256",
    year = "2017"
}

@article{Xing:2020ijf,
    author = "Xing, Zhi-zhong",
    title = "{Flavor structures of charged fermions and massive neutrinos}",
    eprint = "1909.09610",
    archivePrefix = "arXiv",
    primaryClass = "hep-ph",
    doi = "10.1016/j.physrep.2020.02.001",
    journal = "Phys. Rept.",
    volume = "854",
    pages = "1--147",
    year = "2020"
}

@article{Wu:2022tpe,
    author = "Wu, Yongcheng and Xie, Ke-Pan and Zhou, Ye-Ling",
    title = "{Classification of Abelian domain walls}",
    eprint = "2205.11529",
    archivePrefix = "arXiv",
    primaryClass = "hep-ph",
    doi = "10.1103/PhysRevD.106.075019",
    journal = "Phys. Rev. D",
    volume = "106",
    number = "7",
    pages = "075019",
    year = "2022"
}

@article{Fu:2024jhu,
    author = "Fu, Bowen and King, Stephen F. and Marsili, Luca and Pascoli, Silvia and Turner, Jessica and Zhou, Ye-Ling",
    title = "{Non-Abelian Domain Walls and Gravitational Waves}",
    eprint = "2409.16359",
    archivePrefix = "arXiv",
    primaryClass = "hep-ph",
    reportNumber = "IPPP/24/62",
    month = "9",
    year = "2024"
}

@article{Haixing2017,
  title = {Towards the Fundamental Quantum Limit of Linear Measurements of Classical Signals},
  author = {Miao, Haixing and Adhikari, Rana X and Ma, Yiqiu and Pang, Belinda and Chen, Yanbei},
  journal = {Phys. Rev. Lett.},
  volume = {119},
  issue = {5},
  pages = {050801},
  numpages = {6},
  year = {2017},
  month = {Aug},
  publisher = {American Physical Society},
  doi = {10.1103/PhysRevLett.119.050801},
  url = {https://link.aps.org/doi/10.1103/PhysRevLett.119.050801}
}

@article{L-shaped,
  title = {Gravitational-Wave Detector for Postmerger Neutron Stars: Beyond the Quantum Loss Limit of the Fabry-Perot-Michelson Interferometer},
  author = {Zhang, Teng and Yang, Huan and Martynov, Denis and Schmidt, Patricia and Miao, Haixing},
  journal = {Phys. Rev. X},
  volume = {13},
  issue = {2},
  pages = {021019},
  numpages = {13},
  year = {2023},
  month = {May},
  publisher = {American Physical Society},
  doi = {10.1103/PhysRevX.13.021019},
  url = {https://link.aps.org/doi/10.1103/PhysRevX.13.021019}
}

@article{yanbeichen,
doi = {10.1088/0953-4075/46/10/104001},
url = {https://dx.doi.org/10.1088/0953-4075/46/10/104001},
year = {2013},
month = {may},
publisher = {IOP Publishing},
volume = {46},
number = {10},
pages = {104001},
author = {Yanbei Chen},
title = {Macroscopic quantum mechanics: theory and experimental concepts of optomechanics},
journal = {Journal of Physics B: Atomic, Molecular and Optical Physics}
}

@article{Espinosa:2010hh,
    author = "Espinosa, Jose R. and Konstandin, Thomas and No, Jose M. and Servant, Geraldine",
    title = "{Energy Budget of Cosmological First-order Phase Transitions}",
    eprint = "1004.4187",
    archivePrefix = "arXiv",
    primaryClass = "hep-ph",
    reportNumber = "CERN-PH-TH-2010-027",
    doi = "10.1088/1475-7516/2010/06/028",
    journal = "JCAP",
    volume = "06",
    pages = "028",
    year = "2010"
}

@article{Weir:2017wfa,
    author = "Weir, David J.",
    title = "{Gravitational waves from a first order electroweak phase transition: a brief review}",
    eprint = "1705.01783",
    archivePrefix = "arXiv",
    primaryClass = "hep-ph",
    reportNumber = "HIP-2017-06-TH, HIP-2017-06/TH",
    doi = "10.1098/rsta.2017.0126",
    journal = "Phil. Trans. Roy. Soc. Lond. A",
    volume = "376",
    number = "2114",
    pages = "20170126",
    year = "2018",
    note = "[Erratum: Phil.Trans.Roy.Soc.Lond.A 381, 20230212 (2023)]"
}

@article{Huang:2020bbe,
    author = "Huang, Wei-Chih and Sannino, Francesco and Wang, Zhi-Wei",
    title = "{Gravitational Waves from Pati-Salam Dynamics}",
    eprint = "2004.02332",
    archivePrefix = "arXiv",
    primaryClass = "hep-ph",
    reportNumber = "CP3-Origins-2020-05 DNRF90",
    doi = "10.1103/PhysRevD.102.095025",
    journal = "Phys. Rev. D",
    volume = "102",
    number = "9",
    pages = "095025",
    year = "2020"
}

@article{Croon:2018kqn,
    author = "Croon, Djuna and Gonzalo, Tom\'as E. and White, Graham",
    title = "{Gravitational Waves from a Pati-Salam Phase Transition}",
    eprint = "1812.02747",
    archivePrefix = "arXiv",
    primaryClass = "hep-ph",
    reportNumber = "CoEPP-MN-18-11, CoEPP-MN-18-10",
    doi = "10.1007/JHEP02(2019)083",
    journal = "JHEP",
    volume = "02",
    pages = "083",
    year = "2019"
}

@misc{dagnolo2024classical,
    title={Classical (and Quantum) Heuristics for Gravitational Wave Detection},
    author={Raffaele Tito D'Agnolo and Sebastian A. R. Ellis},
    year={2024},
    eprint={2412.17897},
    archivePrefix={arXiv},
    primaryClass={gr-qc}
}

@article{Mohapatra:1980qe,
    author = "Mohapatra, Rabindra N. and Marshak, R. E.",
    title = "{Local B-L Symmetry of Electroweak Interactions, Majorana Neutrinos and Neutron Oscillations}",
    reportNumber = "VPI-HEP-80/1",
    doi = "10.1103/PhysRevLett.44.1316",
    journal = "Phys. Rev. Lett.",
    volume = "44",
    pages = "1316--1319",
    year = "1980",
    note = "[Erratum: Phys.Rev.Lett. 44, 1643 (1980)]"
}

@article{Minkowski:1977sc,
    author = "Minkowski, Peter",
    title = "{$\mu \to e\gamma$ at a Rate of One Out of $10^{9}$ Muon Decays?}",
    reportNumber = "Print-77-0182 (BERN)",
    doi = "10.1016/0370-2693(77)90435-X",
    journal = "Phys. Lett. B",
    volume = "67",
    pages = "421--428",
    year = "1977"
}

@article{Yanagida:1980xy,
    author = "Yanagida, Tsutomu",
    title = "{Horizontal Symmetry and Masses of Neutrinos}",
    reportNumber = "TU-80-208",
    doi = "10.1143/PTP.64.1103",
    journal = "Prog. Theor. Phys.",
    volume = "64",
    pages = "1103",
    year = "1980"
}

@article{Vilenkin:1984ib,
    author = "Vilenkin, Alexander",
    title = "{Cosmic Strings and Domain Walls}",
    reportNumber = "PRINT-84-0840 (TUFTS)",
    doi = "10.1016/0370-1573(85)90033-X",
    journal = "Phys. Rept.",
    volume = "121",
    pages = "263--315",
    year = "1985"
}

@article{Blanco-Pillado:2017oxo,
    author = "Blanco-Pillado, Jose J. and Olum, Ken D.",
    title = "{Stochastic gravitational wave background from smoothed cosmic string loops}",
    eprint = "1709.02693",
    archivePrefix = "arXiv",
    primaryClass = "astro-ph.CO",
    doi = "10.1103/PhysRevD.96.104046",
    journal = "Phys. Rev. D",
    volume = "96",
    number = "10",
    pages = "104046",
    year = "2017"
}

@article{Buchmuller:2019gfy,
    author = "Buchmuller, Wilfried and Domcke, Valerie and Murayama, Hitoshi and Schmitz, Kai",
    title = "{Probing the scale of grand unification with gravitational waves}",
    eprint = "1912.03695",
    archivePrefix = "arXiv",
    primaryClass = "hep-ph",
    reportNumber = "CERN-TH-2019-215, DESY-19-210, DESY 19-210, IPMU 19-0179",
    doi = "10.1016/j.physletb.2020.135764",
    journal = "Phys. Lett. B",
    volume = "809",
    pages = "135764",
    year = "2020"
}

@article{Aasi2015c,
   author = {The LIGO Scientific Collaboration},
   doi = {10.1088/0264-9381/32/7/074001},
   issn = {13616382},
   issue = {7},
   journal = {Classical and Quantum Gravity},
   title = {Advanced LIGO},
   volume = {32},
   year = {2015},
}

@article{EMcav,
  title = {Electromagnetic antennas for the resonant detection of the stochastic gravitational wave background},
  author = {Herman, Nicolas and Lehoucq, L\'eonard and F\ifmmode \mbox{\H{u}}\else \H{u}\fi{}zfa, Andr\'e},
  journal = {Phys. Rev. D},
  volume = {108},
  issue = {12},
  pages = {124009},
  numpages = {10},
  year = {2023},
  month = {Dec},
  publisher = {American Physical Society},
  doi = {10.1103/PhysRevD.108.124009},
  url = {https://link.aps.org/doi/10.1103/PhysRevD.108.124009}
}

@article{Haixingpra,
  title = {General quantum constraints on detector noise in continuous linear measurements},
  author = {Miao, Haixing},
  journal = {Phys. Rev. A},
  volume = {95},
  issue = {1},
  pages = {012103},
  numpages = {6},
  year = {2017},
  month = {Jan},
  publisher = {American Physical Society},
  doi = {10.1103/PhysRevA.95.012103},
  url = {https://link.aps.org/doi/10.1103/PhysRevA.95.012103}
}

@article{Levsphere_1,
  title = {Searching for New Physics with a Levitated-Sensor-Based Gravitational-Wave Detector},
  author = {Aggarwal, Nancy and Winstone, George P. and Teo, Mae and Baryakhtar, Masha and Larson, Shane L. and Kalogera, Vicky and Geraci, Andrew A.},
  journal = {Phys. Rev. Lett.},
  volume = {128},
  issue = {11},
  pages = {111101},
  numpages = {7},
  year = {2022},
  month = {Mar},
  publisher = {American Physical Society},
  doi = {10.1103/PhysRevLett.128.111101},
  url = {https://link.aps.org/doi/10.1103/PhysRevLett.128.111101}
}

@article{Levsphere_2,
  title = {Detecting High-Frequency Gravitational Waves with Optically Levitated Sensors},
  author = {Arvanitaki, Asimina and Geraci, Andrew A.},
  journal = {Phys. Rev. Lett.},
  volume = {110},
  issue = {7},
  pages = {071105},
  numpages = {5},
  year = {2013},
  month = {Feb},
  publisher = {American Physical Society},
  doi = {10.1103/PhysRevLett.110.071105},
  url = {https://link.aps.org/doi/10.1103/PhysRevLett.110.071105}
}

@misc{photoncounting,
    title={Single-Photon Signal Sideband Detection for High-Power Michelson Interferometers},
    author={Lee McCuller},
    year={2022},
    eprint={2211.04016},
    archivePrefix={arXiv},
    primaryClass={physics.ins-det}
}

@article{King:2020hyd,
    author = "King, Stephen F. and Pascoli, Silvia and Turner, Jessica and Zhou, Ye-Ling",
    title = "{Gravitational Waves and Proton Decay: Complementary Windows into Grand Unified Theories}",
    eprint = "2005.13549",
    archivePrefix = "arXiv",
    primaryClass = "hep-ph",
    reportNumber = "FERMILAB-PUB-20-187-T, IPPP/20/20",
    doi = "10.1103/PhysRevLett.126.021802",
    journal = "Phys. Rev. Lett.",
    volume = "126",
    number = "2",
    pages = "021802",
    year = "2021"
}

@article{Aulakh:1982sw,
    author = "Aulakh, C. S. and Mohapatra, Rabindra N.",
    title = "{Implications of Supersymmetric SO(10) Grand Unification}",
    reportNumber = "CCNY-HEP-82-4-REV, CCNY-HEP-82-4",
    doi = "10.1103/PhysRevD.28.217",
    journal = "Phys. Rev. D",
    volume = "28",
    pages = "217",
    year = "1983"
}

@article{Gursey:1975ki,
    author = "Gursey, F. and Ramond, Pierre and Sikivie, P.",
    title = "{A Universal Gauge Theory Model Based on E6}",
    reportNumber = "YALE-3075-118",
    doi = "10.1016/0370-2693(76)90417-2",
    journal = "Phys. Lett. B",
    volume = "60",
    pages = "177--180",
    year = "1976"
}

@article{Bertolini:2009qj,
    author = "Bertolini, Stefano and Di Luzio, Luca and Malinsky, Michal",
    title = "{Intermediate mass scales in the non-supersymmetric SO(10) grand unification: A Reappraisal}",
    eprint = "0903.4049",
    archivePrefix = "arXiv",
    primaryClass = "hep-ph",
    doi = "10.1103/PhysRevD.80.015013",
    journal = "Phys. Rev. D",
    volume = "80",
    pages = "015013",
    year = "2009"
}

@article{Chakrabortty:2019fov,
    author = "Chakrabortty, Joydeep and Maji, Rinku and King, Stephen F.",
    title = "{Unification, Proton Decay and Topological Defects in non-SUSY GUTs with Thresholds}",
    eprint = "1901.05867",
    archivePrefix = "arXiv",
    primaryClass = "hep-ph",
    doi = "10.1103/PhysRevD.99.095008",
    journal = "Phys. Rev. D",
    volume = "99",
    number = "9",
    pages = "095008",
    year = "2019"
}

@article{Lazarides:1984pq,
    author = "Lazarides, George and Shafi, Q.",
    title = "{Extended Structures at Intermediate Scales in an Inflationary Cosmology}",
    reportNumber = "UT-STPD-3/84, BA-84-24",
    doi = "10.1016/0370-2693(84)91605-8",
    journal = "Phys. Lett. B",
    volume = "148",
    pages = "35--38",
    year = "1984"
}

@article{Kephart:2001ix,
    author = "Kephart, Thomas W. and Shafi, Qaisar",
    title = "{Family unification, exotic states and magnetic monopoles}",
    eprint = "hep-ph/0105237",
    archivePrefix = "arXiv",
    reportNumber = "VAND-TH-01-4, BA-01-18",
    doi = "10.1016/S0370-2693(01)01187-X",
    journal = "Phys. Lett. B",
    volume = "520",
    pages = "313--316",
    year = "2001"
}

@article{prd17,
  title = {Frequency-dependent responses in third generation gravitational-wave detectors},
  author = {Essick, Reed and Vitale, Salvatore and Evans, Matthew},
  journal = {Phys. Rev. D},
  volume = {96},
  issue = {8},
  pages = {084004},
  numpages = {7},
  year = {2017},
  month = {Oct},
  publisher = {American Physical Society},
  doi = {10.1103/PhysRevD.96.084004},
  url = {https://link.aps.org/doi/10.1103/PhysRevD.96.084004}
}

@misc{monras2013phase,
    title={Phase space formalism for quantum estimation of Gaussian states},
    author={Alex Monras},
    year={2013},
    eprint={1303.3682},
    archivePrefix={arXiv},
    primaryClass={quant-ph}
}

@article{KLMTV,
  title = {Conversion of conventional gravitational-wave interferometers into quantum nondemolition interferometers by modifying their input and/or output optics},
  author = {Kimble, H. J. and Levin, Yuri and Matsko, Andrey B. and Thorne, Kip S. and Vyatchanin, Sergey P.},
  journal = {Phys. Rev. D},
  volume = {65},
  issue = {2},
  pages = {022002},
  numpages = {31},
  year = {2001},
  month = {Dec},
  publisher = {American Physical Society},
  doi = {10.1103/PhysRevD.65.022002},
  url = {https://link.aps.org/doi/10.1103/PhysRevD.65.022002}
}

@article{Caves,
  title = {On the measurement of a weak classical force coupled to a quantum-mechanical oscillator. I. Issues of principle},
  author = {Caves, Carlton M. and Thorne, Kip S. and Drever, Ronald W. P. and Sandberg, Vernon D. and Zimmermann, Mark},
  journal = {Rev. Mod. Phys.},
  volume = {52},
  issue = {2},
  pages = {341--392},
  numpages = {0},
  year = {1980},
  month = {Apr},
  publisher = {American Physical Society},
  doi = {10.1103/RevModPhys.52.341},
  url = {https://link.aps.org/doi/10.1103/RevModPhys.52.341}
}

@article{EPR,
Author = {Ma, Yiqiu and Miao, Haixing and Pang, Belinda Heyun and Evans, Matthew
   and Zhao, Chunnong and Harms, Jan and Schnabel, Roman and Chen, Yanbei},
Title = {Proposal for gravitational-wave detection beyond the standard quantum
   limit through EPR entanglement},
Journal = {NATURE PHYSICS},
Year = {2017},
Volume = {13},
Number = {8},
Pages = {776-780},
Month = {AUG},
DOI = {10.1038/NPHYS4118},
ISSN = {1745-2473},
EISSN = {1745-2481},
ResearcherID-Numbers = {Miao, Haixing/O-1300-2013
   Schnabel, Roman/V-7759-2019
   Ma, Yiqiu/ISA-4281-2023
   Chen, Yanbei/A-2604-2013
   Zhao, Chunnong/C-2403-2013
   Harms, Jan/J-4359-2012
   },
ORCID-Numbers = {Zhao, Chunnong/0000-0001-5825-2401
   Harms, Jan/0000-0002-7332-9806
   Evans, Matthew/0000-0001-8459-4499
   Miao, Haixing/0000-0003-2879-5821},
Unique-ID = {WOS:000406778100017},
}

@article{cross_lin,
  title = {Sensitivity curves for searches for gravitational-wave backgrounds},
  author = {Thrane, Eric and Romano, Joseph D.},
  journal = {Phys. Rev. D},
  volume = {88},
  issue = {12},
  pages = {124032},
  numpages = {11},
  year = {2013},
  month = {Dec},
  publisher = {American Physical Society},
  doi = {10.1103/PhysRevD.88.124032},
  url = {https://link.aps.org/doi/10.1103/PhysRevD.88.124032}
}

@article{EQL1,
  author  = {F. Ya. Khalili},
  title   = {On the Ultimate Sensitivity of Quantum Probe Systems},
  journal = {Vestnik Moskovskovo Universiteta, Seriya 3: Fizika, Astronomiya},
  number  = {3},
  pages   = {17},
  year    = {1983},
  language = {Russian}
}

@ARTICLE{EQL2,
       author = {{Braginski{\v{i}}}, V.~B. and {Vorontsov}, Yu. I. and {Khalili}, F. Ya.},
        title = "{Quantum singularities of a ponderomotive meter of electromagnetic energy}",
      journal = {Soviet Journal of Experimental and Theoretical Physics},
         year = 1977,
        month = oct,
       volume = {46},
        pages = {705},
       adsurl = {https://ui.adsabs.harvard.edu/abs/1977JETP...46..705B}
}

@inbook{EQL3, 
place={Cambridge}, 
title={Frontmatter}, 
booktitle={Quantum Measurement}, publisher={Cambridge University Press}, author={Braginsky, Vladimir B. and Khalili, Farid Ya and Thorne, Kip S.}, year={1992}, 
pages={i–vi}}

@inproceedings{EQL4,
  author       = {Vladimir B. Braginsky and Mikhail L. Gorodetsky and Farid Ya. Khalili and Kip S. Thorne},
  title        = {Energetic Quantum Limit in Large-Scale Interferometers},
  booktitle    = {Gravitational Waves. Third Edoardo Amaldi Conference, Pasadena, California, 12–16 July 1999},
  editor       = {S. Meshkov},
  series       = {AIP Conference Proceedings},
  volume       = {523},
  pages        = {180--189},
  year         = {2000},
  doi          = {10.1063/1.1291855},
  note         = {Originally published online as arXiv:gr-qc/9907057},
}

@article{axion,
  title = {Photon counting for axion interferometry},
  author = {Yu, Haocun and Kwon, Ohkyung and Namburi, Devendra K. and Hadfield, Robert H. and Grote, Hartmut and Martynov, Denis},
  journal = {Phys. Rev. D},
  volume = {109},
  issue = {9},
  pages = {095042},
  numpages = {7},
  year = {2024},
  month = {May},
  publisher = {American Physical Society},
  doi = {10.1103/PhysRevD.109.095042},
  url = {https://link.aps.org/doi/10.1103/PhysRevD.109.095042}
}

@software{GWINC,
       author = {{Rollins}, Jameson Graef and {Hall}, Evan and {Wipf}, Christopher and {McCuller}, Lee},
        title = "{pygwinc: Gravitational Wave Interferometer Noise Calculator}",
 howpublished = {Astrophysics Source Code Library, record ascl:2007.020},
         year = 2020,
        month = jul,
          eid = {ascl:2007.020},
       adsurl = {https://ui.adsabs.harvard.edu/abs/2020ascl.soft07020R}
}
\enlargethispage{0cm}  

\appendix
\renewcommand\appendix{\par
    \setcounter{section}{0}
    \setcounter{subsection}{0}
    \gdef\thesection{ \Alph{section}}}


\section{Energetic quantum limit}
In this part, we focus on the energetic quantum limit of general system, that is, its gain to the signal is bounded by the vacuum energy fluctuation inside, and therefore related to the total stored energy.
We also dynamically model the impact of quantum loss, extending our discussion to encompass a broader range of systems.
\subsection{Proof of the upper bound of gain}\label{sec:gain}

For a general detection system, the GW signal would introduce a displacement in the out-going quantum state  at certain quadrature
($\hat{b}_{\theta}\equiv\hat{b}_1\cos \theta+\hat{b}_2\sin \theta$) \,\cite{Haixing2017}:
\begin{equation}
\begin{aligned}
\hat{b}_{\theta}(t)&=\hat{b}_{\theta}^{(0)}(t)+\int_{-\infty}^{+\infty} \mathrm{d} t^{\prime} G_{\theta}\left(t-t^{\prime}\right) h\left(t^{\prime}\right)\\
\hat{b}_{\theta}(f)&=\hat{b}_{\theta}^{(0)}(f) +G_{\theta}(f)h(f)\,,
\end{aligned}
\end{equation}
where superscript (0) denotes evolution under a free, time-independent Hamiltonian $\hat{H}_0$, and $G_{\theta}$  is a transfer function describing the system's amplification for strain signal at this quadrature. A system's gain could be defined as the maximum strain-to-displacement transfer functions across all the possible quadratures:
\begin{equation}
|G(f)| = \sup_{\theta} |G_{\theta}(f)|\,.
\end{equation}
With a simple form of interaction Hamiltonian $\hat{H}_{\mathrm{int}}={a} h \hat{\cal E}$, the amplification at certain quadrature  is given by:
\begin{equation}
G_{\theta}\left(t-t^{\prime}\right) \equiv\frac{i{a}}{\hbar}\left[\hat{b}_{\theta}^{(0)}(t), \hat{\cal E}^{(0)}\left(t^{\prime}\right)\right] \Theta\left(t-t^{\prime}\right)\,,
\end{equation}
 which directly yields to an uncertainty principle for continuous measurements\cite{Haixing2017,Haixingpra}:
\begin{equation} \label{Heisenberg}
{a}^2(f)\cdot\left[\bar{S}_{b_{\theta}b_{\theta}}(f) \bar{S}_{\cal{E}\cal{E}}(f)-\left|\bar{S}_{b_{\theta}{\cal E}}(f)\right|^2\right] \geq  {\hbar^2}\left|G_{\theta}(f)\right|^2 \,.
\end{equation}
Here, $\bar{S}_{AB}(f)$  {represents the symmetrized single-sided spectral correlation,}  $\bar{S}_{A B}(f) = S_{AB}(f) + S_{BA}(-f)$, where  the unsymmetrized correlation $S_{AB}$ is defined by:
\begin{equation} \label{S}
\operatorname{Tr}\left[\hat{\rho}_{\mathrm{det}} \hat{A}^{(0)}(f) \hat{B}^{(0) \dagger}\left(f^{\prime}\right)\right] \equiv S_{A B}(f) \delta\left(f-f^{\prime}\right). 
\end{equation}
The equality in Eq.\,\ref{Heisenberg} holds for a general Gaussian pure state \( \hat{b}_{\theta}^0 \). Specifically, the special case of vacuum state injection meets the requirement. Maximizing both side of this equation directly gives the exact value of gain:
\begin{equation} \label{G}
|G(f)|^2 = \frac{{a}^2(f)}{\hbar^2 }\sup_{\theta} \left[ \bar{S}^{\rm vac}_{b_{\theta}b_{\theta}}(f)\bar{S}^{\rm vac}_{\cal EE}(f)-|\bar{S}^{\rm vac}_{b_{\theta}{\cal E}}(f)|^2\right]\,.
\end{equation}
To bound the gain factor, we assume a general input-output relationship and linear form of the cavity mode energy: 
\begin{equation}
\begin{aligned}
\binom{\hat{b}_1(f)}{\hat{b}_2(f)}&=R(\beta_2)\left(\begin{array}{ll}
e^r & 0\\
0& e^{-r}
\end{array}\right) R(\beta_1)\binom{\hat{a}_1(f)}{\hat{a}_2 (f)} \\
 \hat{\cal E}{(f)}&=E_0 \times\left(\cos \varphi, \sin \varphi e^{i \Delta \psi}\right)\binom{\hat{a}_1(f)}{a_2(f)}\,,
\end{aligned}
\end{equation}
For simplicity, we omit the superscript \( (0) \) and an overall phase factor of each.
Here, $\hat{a}_1(f)=\frac{\hat{a}(f)+\hat{a}^{\dagger}(-f)}{\sqrt{2}}$, $\hat{a}_2(f)=\frac{\hat{a}(f)-\hat{a}^{\dagger}(-f)}{\sqrt{2}i}$ is the amplitude and phase quadrature of in-going field respectively, satisfying $\bar{S}_{{a}_1{a}_1}=\bar{S}_{{a}_2{a}_2}=1,\bar{S}_{{a}_1{a}_2}=0$. For the input-output relation, $e^r$ describes the level of internal squeezing, with $R(\beta_1)$ and $R(\beta_2)$ represents the rotation of state before and after the internal squeezing respectively. Meanwhile, $ \varphi$ and $\psi$ basically describes the detector's dynamical response to external field.
With this notation, each term on the right-hand side of Eq.\,\ref{G} can be expanded into the following form:

\begin{small}
\begin{equation}
\begin{aligned}
 \bar{S}_{b_{\theta}b_{\theta}}(f)&=\left[e^r \cos \beta_1 \cos (\theta-\beta_2)+e^{-r} \sin \beta_1 \sin (\theta-\beta_2)\right]^2 \\
& +\left[-e^{r} \sin\beta_1 \cos (\theta-\beta_2)+e^{-r} \cos \beta_1 \sin(\theta-\beta_2)\right]^2 \\
 \bar{S}_{\cal EE}^{v a c}(f)&=E_0^2 \\
 \left|\hat{S}_{b_{\theta}{\cal E}}^{\operatorname{vac}}(f)\right|^2&=E_0^2\mid \cos \varphi\left[e^r \cos \beta_1 \cos (\theta-\beta_2)\right.\\
&+\left.e^{-r} \sin \beta_1 \sin (\theta-\beta_2)\right]+\sin \varphi e^{i \Delta\psi}\times\\
\left[\right.&\left.-e^{{r}} \sin \beta_1 \cos (\theta-\beta_2)+e^{-r} \cos \beta_1 \sin (\theta-\beta_2)\right] \left.\right|^2 \,.
\end{aligned}
\end{equation}
\end{small}
Maximizing  over $\theta$ gives the analytical expression of gain:
\begin{equation}\label{gainbound}
\begin{aligned}
 |G(f)|^2 = \frac{{a}^2\ \bar{S}^{\rm vac}_{\cal EE}(f) }{\hbar^2}\left[ A+\sqrt{A^2-B^2}\right]\,,
\end{aligned}
\end{equation}
where 
\begin{equation}
\begin{aligned}
 A&=\frac{1}{2}\{\cosh (2 r)+[\sin 2 \beta_1 \sin 2 \varphi \cos \Delta \psi \\
&-\cos 2 \beta_1 \cos 2 \varphi] \sinh (2 r)\} \\
 B&=\frac{1}{2} \sin 2 \varphi \sin \Delta \psi \,.
\end{aligned}
\end{equation}

For ultra-high frequency detection, the mechanical back action is usually negligible, corresponding to weak internal squeezing cases  with \( e^{2r} \approx 1 \). In this scenario, Eq.\,\ref{gainbound} could be further simplified as:

\begin{equation} \label{bound}
\begin{aligned}
|G(f)|^2 = \frac{{a}^2 \bar{S}^{\rm vac}_{\cal EE}(f)}{2\hbar^2} \left[ 1+\sqrt{1-\sin^22\varphi\sin^2\Delta\psi}\right]\,.
\end{aligned}
\end{equation}
When \( \Delta \psi = 0 \), which indicates that the energy fluctuation at the upper and lower sidebands has the same magnitude and physically corresponds to the cavity resonant modes being symmetrically distributed around the carrier, a scenario that encompasses all cases discussed in the main text, Eq.\,\ref{bound} directly yields Eq.\,3 in the main text.

Meanwhile, in more general cases, the system's gain can also be bounded by Eq.\,\ref{gainbound}. For instance, in the weak-internal-squeezing scenario with \( \Delta\psi \neq 0 \), the dynamics of the detection system can influence the gain by a factor of up to 2:

\begin{equation} \label{generalbound}
\begin{aligned}
\frac{{a}^2\bar{S}^{\rm vac}_{\cal EE}(f)}{2\hbar^2}\leq|G(f)|^2 \leq \frac{{a}^2\bar{S}^{\rm vac}_{\cal EE}(f)}{\hbar^2}\,.
\end{aligned}
\end{equation}

More generally, for system  with strong dynamical back action (\( e^{2r} > 1 \)), the gain is constrained by:
\begin{equation} \label{generalbound2}
\begin{aligned}
|G(f)|^2 \leq \frac{{a}^2\bar{S}^{\rm vac}_{\cal EE}(f)}{\hbar^2}e^{2r}\,,
\end{aligned}
\end{equation}

\subsection{Derivation of energy fluctuation(Eq.\,\ref{eq:lineshape})} \label{sec:energy}
\textcolor{black}{Since optical gain is bounded by the vacuum energy fluctuation inside the detection system, accurately estimating the energy fluctuation of cavity modes is essential. Here, we derive the Lorentzian lineshape of the energy fluctuation (Eq.\,(4) in the main text).}

We model a general ultra-high-frequency detector as consisting of resonant cavity modes that amplify gravitational wave (GW) signals and external continuous modes to extract information from the cavity. Mechanical back-action, typically weak at these frequencies, is neglected here. {We firstly consider a single, non-degenerate cavity mode that perfectly couples with external continuous field.} The Hamiltonian can be expressed as \cite{yanbeichen}:
\begin{equation}\label{Hamitonian}
\begin{aligned}
\hat{H}&= \hat{\cal E}^{(0)}  +\int_0^{+\infty} 2\pi f\hbar \hat{c}^{\dagger}(f)\hat{c}(f)df 
\\&+i\hbar\sqrt{2\gamma}\int_0^{+\infty} df \left[\hat{A}_{\rm tot}^{\dagger}\hat{c}(f)-\hat{A}_{\rm tot}\hat{c}^{\dagger}(f)\right]+a h\hat{\cal E}^{(0)}\,,
\end{aligned}
\end{equation}
where $\hat{\cal E}^{(0)}=\hbar(\omega_0+\Delta)\hat{A}_{\rm tot}^{\dagger}\hat{A}_{\rm tot}$ represents the total cavity-mode energy, with $\hat{A}_{\rm tot}$ the annihilation operator, and $\Delta>0$ is the detuning between carrier frequency and cavity resonance. The second, third, fourth term represents the Hamiltonian of continuous external mode, the interaction between external field and cavity mode, and the interaction between cavity mode and GW, respectively. Here, the annihilation operator of external field is denoted as $\hat{c}(f)$, while $\omega_0$ and $h$ represent the carrier angular frequency and gravitational-wave amplitude , respectively. The coupling coefficient $a$ relies on both the frequency and the incidence angle of GW signal. It's exact definition is given by:
\begin{equation}
    {a} h \to {\int h^2(f) \overline{a^2}(f)e^{2\pi ift}df}\,,
\end{equation}
where the overline represents the spatial average over all the incidence angle.

To linearize the dynamics, we expand the annihilation operator around its steady-state amplitude:
\begin{equation}
    \hat{A}_{\rm tot} = \bar{A} + \hat{A}\,,
\end{equation}
where $\bar{A}$ corresponds to the classical steady-state amplitude related to stored energy via $\hbar(\omega_0+\Delta)\bar{A}^2=\bar{\cal E}$, and $\hat{A}$ is the quantum fluctuation operator. The linearized cavity energy thus reads:
\begin{equation}\label{E0}
    \hat{\cal E}^{(0)} =\bar{\cal E}+\sqrt{\hbar(\omega_0+\Delta)\bar{\cal E}}(\hat{A}+\hat{A}^{\dagger})\,.
\end{equation}

Applying the linearized approximation, the Heisenberg equations for the cavity fluctuations are:
\begin{equation}\label{model}
\begin{aligned}
-i\omega\hat{A}_{\omega}&=(-i\Delta-\gamma)\hat{A}_{\omega}+\sqrt{2\gamma}\hat{a}_{\omega_0+\omega}+\bar{A}{a}(f)h(f)\\
-i\omega\hat{A}_{\omega}^{\dagger}&=(i\Delta-\gamma)\hat{A}_{\omega}^{\dagger}+\sqrt{2\gamma}\hat{a}_{\omega_0-\omega}-\bar{A}{a}(f)h(f)\,,
\end{aligned}
\end{equation}
where $\hat{a}_{\omega_0+\omega}$ denotes the input field, and $\hat{A}_{\omega}$ is the cavity mode spectrum. Solving Eq.\,\eqref{model}, we have:
\begin{equation}\label{cavityresult}
\hat{A}_{\omega}^{(0)}+\hat{A}_{\omega}^{\dagger(0)}=\frac{2\sqrt{\gamma}\left[(2\pi f-i\gamma)\hat{a}_{1\omega}-i\Delta\hat{a}_{2\omega}\right]}{(2\pi f-i\gamma)^2-\Delta^2}\,,
\end{equation}
with amplitude and phase quadratures defined as $\hat{a}_{1\omega}=(\hat{a}_{\omega_0+\omega}+\hat{a}^{\dagger}_{\omega_0-\omega})/\sqrt{2}$ and $\hat{a}_{2\omega}=(\hat{a}_{\omega_0+\omega}-\hat{a}^{\dagger}_{\omega_0-\omega})/(\sqrt{2}i)$.
Combining Eqs.\,\eqref{E0} and \eqref{model}, and using vacuum-state correlations, the cavity-mode energy fluctuation is given by:
\begin{equation}\label{SEE}
S^{\rm vac}_{\cal EE}(f)=\hbar(\omega_0+\Delta)\bar{\cal E}\cdot\frac{4\gamma^2\left[(2\pi f)^2+\gamma^2+\Delta^2\right]}{\left[(2\pi f)^2+\gamma^2-\Delta^2\right]^2}\,.
\end{equation}
Near resonance ($2\pi f\approx\Delta$) and under the condition $\Delta\gg\gamma$, Eq.\,\eqref{SEE} simplifies significantly to a Lorentzian:
\begin{equation}\label{Shh2}
S^{\rm vac}_{\cal EE}(f)\approx{2\gamma\bar{\cal E}\hbar}\frac{\omega_0+\Delta}{(2\pi f-\Delta)^2+\gamma^2}\,.
\end{equation}
The corresponding single-sided symmetrized correlation is thus:
\begin{equation}
\bar{S}^{\rm vac}_{\cal EE}(f)\approx 2\gamma\bar{\cal E}\hbar\frac{\omega_0+\Delta}{(2\pi f-|\Delta|)^2+\gamma^2}={\hbar}\bar{\cal E}S(f)\,,
\end{equation}
with $S(f)$ being a normalized Lorentzian lineshape consistent with Eq.\,(4) in the main text.

For detectors with multiple {cavity} resonant modes, the total energy fluctuation is the sum of individual Lorentzian contributions:
\begin{equation}\label{Shhmulti}
\begin{aligned}
\bar{S}^{\rm vac}_{\cal EE}(f)&\approx\sum_k \sum_{i=1}^{g_k} \,|\eta_{ki}|^2\cdot\frac{2\gamma_k\bar{\cal E}\hbar(\omega_0+\Delta_k)}{(2\pi f-|\Delta_k|)^2+\gamma_k^2}\\
&=\sum_k \frac{2\gamma_k\bar{\cal E}\hbar(\omega_0+\Delta_k)}{(2\pi f-|\Delta_k|)^2+\gamma_k^2}\,,
\end{aligned}
\end{equation}
{where the subscript $k$ indexes the distinct resonant frequencies, 
and $i$ labels the degenerate states associated with the $k$-th frequency, 
with $g_k$ denoting the corresponding degree of degeneracy. 
The dimensionless scalar $\eta_{ki}$ denotes the coupling strength 
between the external mode and the $\{k,i\}$-th state , satisfying the 
normalization condition:
$\sum_{i=1}^{g_k} |\eta_{ki}|^2 = 1.$
}{The result} in align with Eq.\,(4) in the main text.

With the discussion above, we could also give a uniform estimation of the peak sensitivity of general detector. Near the cavity resonance  \( f = \frac{\Delta_k}{2\pi} \), the system's gain to GW signal is primarily contributed by this mode, which could be estimated as:
\begin{equation}\label{Gpeak}
\begin{aligned}
|G_{\rm peak}|^2 &=\frac{\bar{S}_{\cal EE}(\frac{\Delta_k}{2\pi}) \overline{{a}^2}(\frac{\Delta_k}{2\pi})}{\hbar^2} \\
&\sim \frac{2\bar{\cal E}(\omega_0+\Delta_k)\overline{{a}^2}(f_0)}{\hbar \gamma } = \frac{2\bar{\cal E}Q_k \overline{{a}^2}(f_0)}{\hbar}\,,
\end{aligned}
\end{equation}
where $Q_k=\frac{\omega_0+\Delta_k}{\gamma}$ is the quality factor of this cavity mode. 
Intuitively, to achieve large gain for gravitational wave signals, a detector must have a sufficiently large energy storage and a well-calibrated resonant mode with narrow bandwidth. This conclusion provides a direct intuition for the future design of detectors.

\subsection{Model of quantum loss} \label{sec:loss}

The previous discussion primarily focuses on the ideal scenario without quantum loss.
In realistic situations, interactions between the quantum state and the environment introduce additional noise, referred to as quantum loss. {Here, we focus specifically on the contribution from cold quantum loss. With this type of quantum loss, the quantum state is partially scattered to thermal vacuum state while interacting with the detector, modeled by the following Hamiltonian:}

\begin{equation}\label{Hamitonian_loss}
\begin{aligned}
\hat{H}_{\rm loss,{k}}&= 
i \hbar \sqrt{2 \gamma {\cal L}_k}\int_0^{+\infty}\left[\hat{A}_{\rm tot}^{\dagger} \hat{n}^0_k(f)-\hat{A}_{\rm tot} \hat{n}_k^{0\dagger}(f)\right] df\\
&\quad+\int_0^{+\infty} 2\pi f\hbar\,\hat{n}_k^{0\dagger}(f)\hat{n}^0_k(f)\,df\,,
\end{aligned}
\end{equation}
where the label $k$ represents different sources of quantum loss,  ${\cal L}_k{\ll1}$ {is a unit-less parameter characterizing the relative coupling strength of the loss channel to the cavity mode}, and $\gamma$ represents the linewidth of the cavity mode. Here, $\hat{n}^0_k$ are vacuum states with $\bar{S}_{\hat{n}^0_k\hat{n}^0_k}=1$ .
{For simplicity, we only consider non-degenerate cavity mode here. Systems with degenerate cavity modes basically follows similar mechanism.}

When coupled to several loss channels, the dynamical evolution of cavity mode would be modified as follows:
\begin{equation} \label{model_lin}
\begin{aligned}
-i \omega \hat{A}_{\omega}&=(-i \Delta-\gamma)\hat{A}_{\omega}+\sqrt{2 \gamma}\hat{a}_{\omega_0+\omega}+\bar{A}\left[{a} h+\sum_k\alpha_k\hat{n}_k^0\right]\,, \\
-i \omega \hat{A}_{\omega}^{\dagger}&=(i \Delta-\gamma)\hat{A}_{\omega}^{\dagger}+\sqrt{2 \gamma}\hat{a}_{\omega_0-\omega}-\bar{A}\left[{a} h+\sum_k\alpha_k\hat{n}_k^0\right]\,,
\end{aligned}
\end{equation}
where {we assume ${\cal L}_k \ll 1$}. Here,  the {strength of dissipation $\alpha_k(f)$ is given by:}
{
\begin{equation}\label{Lossrate}
|\alpha_k|^2=\frac{2{\cal L}_k \gamma \hbar(\omega_0+\Delta)}{\bar{\cal E}}\,,
\end{equation}}
{where $\omega_0$ is the angular frequency of carrier $, \Delta$ is the detuning, and $\gamma$ is the bandwidth of cavity mode }. {Physically, the admixture of thermal states $\hat{n}_k^0$ introduces an additional broadband fluctuation background that is inseparably attached to the strain signal, and cannot be eliminated by any quantum operation or readout strategy on the continuous mode $\hat{a}_{\omega}$.}

In the presence of quantum loss, the output field acquires a thermal component, modifying the ideal input-output relation to:
\begin{equation}\label{output}
\hat{b}_{\theta} = \left[1-\sum_k\varepsilon_k(f)\right]^{{1/2}}\hat{b}_{\theta}^0 +G_{\theta}(f)h(f)+\sum_k\sqrt{\varepsilon_k(f)}\hat{n}_{k{,\theta}}\,,
\end{equation}
where $\hat{b}_{\theta}^0$ denotes the ideal output state without quantum loss, , \( \hat{n}_k \) represents the thermal state component introduced by different sources at the output, and $\varepsilon_k(f)$ represents the {scattering probability} into the $k$-th loss channel. The total probability of decoherence, characterizing the quantum loss level of the system effectively, is connected to the detector’s intrinsic dynamics by:
\begin{equation}\label{loss_final}
\varepsilon(f)=\sum_k\varepsilon_k(f)={\sum_k \left[\frac{\bar{S}_{\cal EE}^{\rm vac}(f)\cdot 2{\cal L}_k \gamma}{\hbar \bar{\cal E}(\omega_0+\Delta)}\right]}\,.
\end{equation}

{
For detectors with internal squeezing, the gain can exceed the energetic quantum limit by a factor up to $e^{2r}$, and the loss components $\hat{n}_{k}$ are no longer in a vacuum state at the readout, but instead correspond to squeezed vacuum states with
$S_{\hat{n}_{k,\theta},\,\hat{n}_{k,\theta}} \in [e^{-2r},\,e^{2r}]$.
Nevertheless, for those loss channels that are induced before the internal squeezing, the fundamental indistinguishability between the strain signal and quantum loss remains unchanged.  Specifically, for linear measurement that tracks the strain signal $h$, the fundamental loss-limited noise background is bounded by:
}
{
\begin{equation}
S_{hh}^{\rm noise}=\frac{S_{\hat{b}_\theta\hat{b}_\theta}}{|G_\theta|^2 }\geq  \frac{\sum_k \varepsilon_k S_{\hat{n}_k\hat{n}_k}}{|G_\theta|^2} \geq \frac{\sum_k|\alpha_k|^2}{a^2}= \frac{\varepsilon_b \hbar^2}{a^2\bar{S}_{\cal EE}^{\rm vac}}\,,
\end{equation}}
{where the summation only covers loss channels that are introduced before the internal squeezing. Here, the rightmost term corresponds exactly to the contribution of inseparable noise background appearing in eq.~\ref{model_lin}, which is directly attached to the signal. This loss limit remains for linear measurement remains valid even in the presence of internal squeezing.}

{For detection strategies that directly measure $h^2(t)$ (or equivalently, $S_{hh}(f)$ in frequency domain), Eq.~\ref{model_lin} can also provide a valid sensitivity upper limit:}
{
\begin{equation} \label{eq:sqz}
\begin{aligned}
{\rm SNR}^2 &\le
\frac{\left|\left\langle\left(
a^* h^* + \sum_k \alpha_k^*\hat{n}_k^{0\dagger}\right)
\left(a h + \sum_k \alpha_k \hat{n}_k^{0}\right)\right\rangle\right|^2}
{{\rm Var}\!\left[\left(
a^* h^* + \sum_k \alpha_k^*\hat{n}_k^{0\dagger}\right)
\left(a h + \sum_k \alpha_k \hat{n}_k^{0}
\right)\right]}\\ &=\frac{(S_{hh} (f))^2}{S_{hh}(f)+\frac{\varepsilon_b(f) \hbar^2}{a^2 \bar{S}_{\cal EE}^{\rm vac}(f)}}\,.
\end{aligned}
\end{equation}
}
{This limit also holds for general cases with internal squeezing.}
\section{Review of Different Measurement Schemes}\label{sec:QFI}

In addition to the optical gain, the measurement scheme significantly impacts the precision of estimating stochastic gravitational wave signals. The effectiveness of a measurement scheme is primarily determined by two factors:

\begin{itemize}
\item[(a)] The initial quantum state, determining the amount of information encoded for a given GW-induced shift.
\item[(b)] The readout method, determining the effectiveness of information extraction from the quantum state.
\end{itemize}

The measurement schemes listed in Table.\,1 fall into three categories: linear measurement, photon counting, and channel limit, whose detailed performance is discussed in \cite{Stochastic}. Here we briefly review their principles and performance, and summarize the derivation of the corresponding SNR limits and minimum detectable thresholds.

\subsection{Review of Fisher Information}
Before examining specific schemes, we briefly introduce the Fisher information, an essential tool to quantitatively describe information encoded in quantum states.

For a parameter $\varphi$ encoded in a quantum state $\hat{\rho}(\varphi)$, the Mean Square Error (MSE) of an unbiased estimation via a measurement with probability distribution \( p(x|\varphi) \) is bounded by the Classical Cramer-Rao Bound (CCRB):
\[
\Delta^2 \varphi \geq \frac{1}{N}\left[\mathcal{F}_C(\varphi)\right]^{-1},
\]
where the Classical Fisher Information (CFI) $\mathcal{F}_C(\varphi)$ is defined as:
\[
\mathcal{F}_C(\varphi)=\int_{-\infty}^{\infty} dx\, \frac{\left[\partial_\varphi p(x|\varphi)\right]^2}{p(x|\varphi)}\,.
\]
The CCRB implies an upper bound on the signal-to-noise ratio (SNR) for parameter estimation:
\begin{equation}
{\rm SNR}[\varphi] \leq \sqrt{N \mathcal{F}_C(\varphi)}\,\bar{\varphi}\,,
\end{equation}
where \( \bar{\varphi} \) is the expectation value of the parameter.

The Quantum Fisher Information (QFI), which bounds the extractable information over all possible measurements, is defined as:
\begin{equation}
  \mathcal{F}_Q(\varphi) = \sup \mathcal{F}_C(\varphi)\,.
\end{equation}
The QFI establishes the Quantum Cramer-Rao Bound (QCRB) on SNR for optimal measurements:
\begin{equation}\label{QCRB_def}
{\rm SNR}[\varphi] \leq \sqrt{N \mathcal{F}_C(\varphi)}\,\bar{\varphi} \leq \sqrt{N \mathcal{F}_Q[\varphi, \hat{\rho}_{\varphi}]}\,\bar{\varphi}\,.
\end{equation}

For a quantum state \( \hat{\rho}_{\varphi} = \sum_j p_j |\phi_j\rangle \langle\phi_j| \) with eigenvalues \( p_j \) and eigenvectors \( |\phi_j\rangle \), the QFI is explicitly:
\[
\mathcal{F}_Q[\varphi, \hat{\rho}_{\varphi}] = \sum_{j,k} \frac{2}{p_j + p_k} \left| \langle \phi_j | \partial_\varphi \hat{\rho}_{\varphi} | \phi_k \rangle \right|^2,\quad (p_j+p_k>0)\,.
\]

Meanwhile, in the simple case where the signal is encoded only in the covariance matrix $\Sigma$ of a Gaussian state, the QFI can be analytically expressed as \cite{monras2013phase}:
\begin{equation} \label{eqn}
\mathcal{F}_Q(\varphi)=\frac{\operatorname{Tr}\left[\left(\Sigma^{-1} \partial_\varphi \Sigma\right)^2\right]}{\left(1+\gamma^2\right)}+\frac{4\left(\partial_\varphi \gamma\right)^2}{1-\gamma^4}\,,
\end{equation}
where \(\gamma=\operatorname{det}(\Sigma/2)^{-\frac{1}{2}}\) is the state’s purity. Here, the covariance matrix is defined in the single-sided formalism.

In our scenario, the goal is to estimate the gravitational-wave-induced fluctuation variance:
\begin{equation}
\sigma^2(f):=|G(f)|^2\overline{{a}^2}(f)h^2(f)\,,
\end{equation}
which fits naturally into this general single-parameter estimation framework. The hierarchical relationship of QFI and QCRB allows systematic assessment of quantum states (via QFI) and readout methods (via saturation of QCRB). Below, we apply this framework to analyze specific measurement schemes listed in Table.\,1.

To avoid divergence of QFI for weak signals, we calculate the QFI with respect to the standard deviation \(\sigma\) instead, which is in aligh with Ref.\,\cite{Stochastic}. The corresponding SNR limit for $\sigma^2$ is given by the chain rule:
\begin{equation}\label{QCRB}
{\rm SNR}\left[\sigma^2\right] \leq \sigma^2\sqrt{N \mathcal{F}_{\mathrm{Q}}\left(\sigma^2\right)} = \frac{\sigma}{2} \sqrt{N \mathcal{F}_{\mathrm{Q}}\left(\sigma\right)}\,.
\end{equation}

\subsection{Linear Measurement}\label{sec:homodyne}
Linear measurement, widely utilized in detectors like LIGO\,\cite{Aasi2015c}, is an effective approach for detecting deterministic gravitational-wave (GW) signals by continuously reading out the GW-induced displacement on quantum states. Here we briefly review its principle and performance.

\vspace{5pt}
\noindent
\textbf{State preparation: Single-Mode Squeezed Vacuum (SMSV)}

\vspace{5pt}

Typically, a Single-Mode Squeezed Vacuum (SMSV) state is chosen as the input state to enhance sensitivity\,\cite{Caves, KLMTV, EPR}. This state suppresses quantum fluctuations in the quadrature encoding the signal, amplifying fluctuations in the orthogonal quadrature. Under weak mechanical back-action and a squeezing factor \( e^{-r} \), the output covariance matrix is:
\begin{equation} \label{covariance_SMSV}
\Sigma_{\rm SMSV}(\sigma)=\left[ \begin{array}{cc}
{(1-\varepsilon)e^{2r}+\varepsilon} & 0 \\
0 & {(1-\varepsilon)e^{-2r}+\varepsilon} + \sigma^2
\end{array} \right]\,,
\end{equation}
where \(\varepsilon\) represents system loss. In the strong squeezing limit (\( r \to \infty \)), applying Eq.\,\ref{eqn}, the Quantum Fisher Information (QFI) is given by:
\begin{equation}
\mathcal{F}_Q(\sigma, {\rm SMSV}) = \frac{4\sigma^2}{(\varepsilon + \sigma^2)^2}\,. 
\end{equation}
In the weak-signal regime (\(\sigma^2\ll\varepsilon\)), the QFI scales as \(\frac{4\sigma^2}{\varepsilon^2}\), and the optimal SNR becomes:
\begin{equation}\label{QCRB_SMSV}   
{\rm SNR}\left[ \sigma^2 \right]=\frac{\sigma^2}{\varepsilon} \sqrt{N}\,.
\end{equation}
With frequency resolution \(\Delta f\) and integration time \(T_{\rm int}\), the corresponding minimum detectable thresholds are:
\begin{equation} \label{hc_hom}
\begin{aligned}
h^{2}_{\rm c,min}(f)&=\frac{\varepsilon f \hbar^2}{ \sqrt{T_{int}\Delta f}\,\bar{S}^{\rm vac}_{\cal EE}(f)\overline{{a}^2}(f)}\,,\\
{\Omega}_{\rm GW,min}(f)&=\frac{4\pi^2}{3H_{0}^2}\frac{\varepsilon f^3 \hbar^2}{ \sqrt{T_{int}\Delta f}\,\bar{S}^{\rm vac}_{\cal EE}(f)\overline{{a}^2}(f)}\,,
\end{aligned}
\end{equation}
which consistent with the results presented in Tab.\,1 of the main text.

The broadband stochastic GW detection SNR limit with SMSV injection is:
\begin{equation}\label{ultimate_SNR_lin_ideal}
\mathrm{SNR}^2_{\rm SMSV} \approx   \int_{f_{min}}^{f_{max}} \left( \frac{h_{\rm c,sig}^2(f)}{h^{2}_{\rm c,min}(f)}\right)^2 \frac{df}{\Delta f}\,.
\end{equation}

\textcolor{black}{Meanwhile, with multiple detectors within identical frequency band, the sensitivity could be further enhanced via cross-correlation.  Optimal joint SNR with k detectors is given by:}
\textcolor{black}{\begin{equation}\label{correlation_lin}
\mathrm{SNR}^2_{\rm SMSV} \approx  k(k-1) \int_{f_{min}}^{f_{max}} {\cal R}_{\rm eff}(f)\left( \frac{h_{\rm c,sig}^2(f)}{h^{2}_{\rm c,min}(f)}\right)^2 \frac{df}{\Delta f}\,,
\end{equation}}
\textcolor{black}{
for co-located detectors, it can bring an $\rm SNR$ improvement of 2 for $k=2$ and $\sim k$  for $k\gg1$. Meanwhile, the order-of-unity overlap reduction function ${\cal R}_{\rm eff}$  depends on both the configuration of individual detector and the global arrangement of detector array. Detailed discussion of overlap reduction function can be found in Ref.\,\cite{cross_lin}.}

\vspace{5pt}
\noindent\textbf{Optimal readout: homodyne detection}

\vspace{5pt}

A standard readout method for SMSV states is homodyne detection, \textcolor{black}{in which a strong classical local oscillator (LO) field 
$E(t)=E_0\cos(\omega_0 t+\theta)$, operating at the same frequency $\omega_0$ as the carrier, is mixed with the output field. 
Same frequency between LO and carrier enables the selection of readout quadrature  by tuning the LO phase $\theta$.} The corresponding optical power spectrum at the readout is then given by:
\begin{equation}\label{power}
\hat{P}(f)={\hbar(\omega_0+\omega)}\bar{b}\cdot\hat{b}_\theta(f)\,,
\end{equation}
where \(\bar{b}\) relates to the mean outgoing power \(\bar{P}(t)\) by \(\hbar \omega_0\bar{b}^2=\bar{P}(t)\), and \(\hat{b}_{\theta}(f)\) is the outgoing quadrature from Eq.\,\ref{output}. Thus, the GW-induced variance is encoded in the power spectral density:
\begin{equation}\label{SPP}
\bar{S}_{PP}(f)= \hbar^2(\omega_0+\omega)^2 \bar{b}^2 \left[\sigma^2(f)+\bar{S}_{b_{\theta_0}b_{\theta_0}}+\sum_k\varepsilon_k\bar{S}_{n_kn_k}\right]\,.
\end{equation}
Detection involves distinguishing the excess variance from a constant quantum noise background.

At high squeezing levels (\(r\to\infty\)), the intrinsic noise \(\bar{S}_{{b}_{\theta_0}{b}_{\theta_0}}\) is negligible, leaving quantum loss as the primary limit. Thus, in practice with weak internal squeezing, the achievable SNR is:
\begin{equation}\label{SNR_hom}
{\rm SNR}_{\rm hom}[\sigma^2]=\frac{\sigma^2}{\sigma^2+\sum_k\varepsilon_k(f)\bar{S}_{n_kn_k}(f)}=\frac{\sigma^2}{\sigma^2+\varepsilon(f)}\,.
\end{equation}
Comparing Eq.\,\ref{SNR_hom} and Eq.\,\ref{QCRB_SMSV} (with \(T_{\rm int}\Delta f=1\)), we confirm that homodyne detection saturates the QCRB for SMSV input, making it optimal. For general cases involving internal squeezing, the optimal SNR is bounded by loss introduced prior to mechanical back-action:
\begin{equation}\label{SNR_hom_backaction}
 \frac{\sigma^2}{\sigma^2+\varepsilon(f)}\leq {\rm SNR}[\sigma^2] \leq \frac{\sigma^2}{\sigma^2+\varepsilon_b(f)}\,,
\end{equation}
where \(\varepsilon_b=\sum_{k_b} \varepsilon_{k_b}\) denotes the quantum loss induced before mechanical back-action.

\textcolor{black}{For the general case with classical noise, the optimal estimator remains unchanged, while the classical noise adds a constant contribution, modifying the optimal SNR to:
\begin{equation}\label{SNR_hom_loss}
{\rm SNR}_{\rm hom}[\sigma^2]=\frac{\sigma^2}{\sigma^2+\varepsilon(f)+\varepsilon_{\rm cl}(f)}\,,
\end{equation}}
\textcolor{black}{with $\varepsilon_{\rm cl}(f)$ denoting the single-sided power spectral density of classical noise normalized to the shot-noise level. The relative magnitude between $\varepsilon_{\rm cl}(f)$ and $\varepsilon(f)$ determines whether a detector can operate at its quantum limit. }

{For laser interferometers, typical sources of classical noise at high frequencies are summarized in Table~\ref{tab:classical}. Suppressing the combined contribution of all classical noise channels below the quantum limit imposes practical requirements on concrete  parameters of detector.}

\begin{table}
\begin{center}
\begin{threeparttable}
\caption{{Typical sources of classical noise at ultra-high frequencies, along with the dominant factors for each noise channel. Detailed mechanisms and exact formulas for different types of noises can be found in the source code of the simulation package, \texttt{pyGWINC}~\cite{GWINC}.} }
\par
\begin{tabular}{cc}
\hline
\textbf{Type} & \textbf{Primary driver(s)} \\
\hline
Coating thermal noise & Loss angle; temperature \\
\hline
Substrate thermal noise & Loss angle; temperature;  \\
(thermal motion of the mirror) &mirror mass \\
\hline
Residual gas noise & Vacuum level; temperature \\
\hline
Laser noise coupling & Stability of source laser ;\\
&optical contrast defect \\
\hline
Readout noise & Photodetectors' performance \\
\hline
\label{tab:classical}
\end{tabular}
\end{threeparttable}
\end{center}
\end{table}
 \vspace{5pt}
\textcolor{black}{\noindent\textbf{Sub-optimal case: heterodyne detection}}

\vspace{5pt}
Another important readout scheme is heterodyne detection, in which the local oscillator is detuned from the carrier by an intermediate frequency $\Omega_{\rm IF}$, i.e.\ $E(t)=E_0\cos[(\omega_0+\Omega_{\rm IF})t+\theta]$. In this case, the output photocurrent contains contributions not only from the signal sideband near $\omega_0$ but also from the image band centered at $\omega_0- \Omega_{\rm IF}$. The.

Formally, the heterodyne power spectrum can be expressed as
\begin{equation}\label{SPP_het}
\begin{aligned}
\bar{S}_{PP}^{\rm het}(f)&= \hbar^2(\omega_0+\omega)^2 \bar{b}^2 \\
&\times\left[\sigma^2(f)+\bar{S}_{b_{\theta_0}b_{\theta_0}}+\sum_k\varepsilon_k\bar{S}_{n_kn_k} 
+ S_{\rm vac}^{\rm img}\right]\,,
\end{aligned}
\end{equation}
where $S_{\rm vac}^{\rm img}$ denotes the irreducible contribution from the image band vacuum, which cannot be suppressed by squeezing injection within the signal band.  The achievable SNR is therefore bounded by the vacuum noise rather than quantum loss
\begin{equation}\label{SNR_het}
{\rm SNR}_{\rm het}[\sigma^2] = 
\frac{\sigma^2}{\sigma^2+\varepsilon(f)+1}\,,
\end{equation}
which is strictly lower than the homodyne case in Eq.\,\ref{SNR_hom}.  

Thus, the performance of heterodyne is intrinsically degraded by unsqueezed image-band noise, and cannot saturate the loss limit achievable with balanced homodyne detection. Nonetheless, the sensitivity upper limit given by fundamental limit also valid in this case. 

\subsection{Photon Counting}\label{}
Recently, a photon-number-preserving measurement scheme known as \textit{photon counting}\,\cite{photoncounting} has been proposed, showing better performance than linear measurement for weak stochastic signals under ideal conditions. Here, we briefly overview its principle and performance.

\vspace{5pt}
\noindent\textbf{State preparation: vacuum state}

\vspace{5pt}

The input state in photon counting is simply the vacuum. When mechanical back-action is negligible, the covariance matrix at the output simplifies to:
\begin{equation} \label{covariance_vac}
\Sigma_{\rm vac}(\sigma)=\left[ \begin{array}{cc}
1 & 0 \\
0 & 1 + \sigma^2
\end{array} \right].
\end{equation}
Using Eq.\,\ref{eqn}, in the weak-signal limit (\(\sigma \ll 1\)), the QFI for the vacuum state is:
\begin{equation}
{\cal F}_Q(\sigma, \rm{vac})=2.   
\end{equation}
Interestingly, for weak stochastic signals (\(\sigma^2 < \varepsilon^2/2\)), the vacuum state exhibits better scaling behavior compared to SMSV, implying that pre-squeezing does not always enhance the distinguishability of stochastic signals.

The optimal achievable SNR using vacuum state injection, given by the Quantum Cramer-Rao Bound (QCRB), reads:
\begin{equation}\label{QCRB_vac}
{\rm SNR}\left[ \sigma^2 \right]=\sqrt{N/2}\,\sigma\,.
\end{equation}
Consequently, the minimum detectable thresholds are:
\begin{equation} \label{hc_pc}
\begin{aligned}
h^2_{\rm c, min}(f)&=\frac{2f\hbar^2}{T_{\rm int}\Delta f \cdot \bar{S}^{\rm vac}_{\cal EE}(f)\overline{{a}^2}(f)},\\
{\Omega}_{\rm GW,min}(f)&=\frac{4\pi^2}{3H_{0}^2}\frac{2f^3\hbar^2}{T_{\rm int}\Delta f \cdot \bar{S}^{\rm vac}_{\cal EE}(f)\overline{{a}^2}(f)}\,,
\end{aligned}
\end{equation}
matching the main text. For broadband signals, the SNR bound  can be expressed as:
\begin{equation}\label{ultimate_SNR_pc_ideal}
\mathrm{SNR}^2_{\rm vac} \approx \int_{f_{min}}^{f_{max}} \frac{h_{\rm c,sig}^2(f)}{h^{2}_{\rm c,min}(f)}\frac{df}{\Delta f}\,.
\end{equation}
{It is worth noting that this sensitivity bound derived for vacuum-state injection applies only to systems without internal squeezing. For systems incorporating internal squeezing, this limit can be surpassed.
 }

\textcolor{black}{The cross-correlation in this quadratic metrology scheme differs significantly from that in linear metrology. Crucially, the signal superposition must be achieved through physical interferometric processes prior to detection, as opposed to computational post-processing of acquired data streams. A detailed discussion of cross-correlation between to co-located detectors with quadratic metrology is given in Ref.\,\cite{photoncounting}, which can  enhance the $\rm SNR$ factor of $\sqrt{2}$.  }

\vspace{5pt}
\noindent\textbf{Optimal readout: photon counting}

\vspace{5pt}

The optimal readout in this scheme is simply photon flux measurement at the dark port in the absence of local oscillator, where the photon flux depends quadratically on the strain signal\,\cite{photoncounting}:
\begin{equation}\label{photonflux}
\dot{n}(t)=\hat{b}^{\dagger}(t)\hat{b}(t)=\hat{b}_0^{\dagger}(t)\hat{b}_0(t)+\int\frac{\sigma^2(f)}{2} df\,.
\end{equation}
Here, \(\hat{b}(t)\) and \(\hat{b}_0(t)\) represent outgoing fields with and without gravitational-wave signals, respectively. Considering only a narrow bandwidth \(\Delta f\) around frequency \( f_0 \), we simplify to:
\begin{equation}
\dot{n}(t) = \frac{\sigma^2(f_0)}{2} \Delta f + \hat{b}_0^{'\dagger}(t)\hat{b}_0^{'}(t)\,,
\end{equation}
with \(\hat{b}_0^{'}(t)\) the filtered intrinsic output field.

Due to extremely low photon flux at the dark port, continuous power measurement is impractical, and the measurable quantity is total photon number \( n_{\rm tot} = T_{\rm int}\dot{n}(t) \) that reach the dart port over integration time \(T_{\rm int}\). The expectation and variance of photon number are:
\begin{equation}
\langle n_{\rm tot} \rangle = \left[\frac{\sigma^2(f_0)}{2}+\frac{\bar{\dot{n}}_0}{\Delta f}\right]T_{\rm int}\Delta f,\quad
\langle \Delta n_{\rm tot}^2 \rangle = \langle n_{\rm tot} \rangle\,,
\end{equation}
where \({\dot{n}}_0= \hat{b}_0^{'\dagger}(t)\hat{b}^{'}_0(t) \) is the intrinsic photon flux without GW signals. when \(\bar{\dot{n}}_0 \to 0\), the measured photon number would obey a Poisson statistics, with the likelihood function and corresponding SNR of:
\begin{equation}
\begin{aligned}
{\cal L}(n_{\rm meas}|n_{\rm tot})
&= \frac{n_{\rm tot}^{\,n_{\rm meas}}}{n_{\rm meas}!}\, e^{-n_{\rm tot}},\,\\
{\rm SNR}^2 = \frac{\langle n_{\rm meas} \rangle^2}{{\rm Var}(n_{\rm meas})}
&\leq \frac{n_{\rm tot}^2}{{\big\langle \big[\partial_{n_{\rm tot}}\ln {{\cal L}(n_{\rm meas}|n_{\rm tot})}\big]^{-2}\big\rangle} }
= n_{\rm tot}\,,
\end{aligned}
\end{equation}
which perfectly saturates the QCRB of vacuum state in Eq.\,\ref{QCRB_vac}.

\textcolor{black}{With non-zero $\bar{\dot{n}}_0$ ,  the optimal estimator for stochastic signal is given by:
\begin{equation}\label{estimator}
\overline{\sigma^2}(f)=\frac{2(n_{\rm tot}-\bar{\dot{n}}_0 T_{\rm int})}{T_{\rm int}\Delta f}\,,
\end{equation}
with an achievable SNR thus is:
\begin{equation}\label{pc_snr}
{\rm SNR}_{\rm pc}[\sigma^2]=\sqrt{T_{\rm int}\Delta f}\frac{\sigma^2}{\sqrt{2(\sigma^2+2\bar{\dot{n}}_0/\Delta f)}}\,,
\end{equation}}
\textcolor{black}{from which we can see the non-zero $\bar{\dot{n}}_0$ would degrade the sensitivity. If $\bar{\dot{n}}_0 \Delta f \lesssim \sigma^2$, photon counting can still exhibit the Poisson scaling of the ideal case; otherwise, its scaling behavior will degrade to that of homodyne detection, with $\rm SNR \propto \sigma^2$.}
\textcolor{black}{In practical scenario, \(\bar{\dot{n}}_0=0\) cannot be strictly maintained. Since the input state is the vacuum, quantum loss does not generate excess noise photons. A non-zero photon flux may arise from the following mechanisms:
\begin{itemize}
\item Dark count, which can be understood as false positives of the single-photon detector;
\item  {Classical noises. Possible sources of classical noise within laser interferometer are summarized in Table.\,\ref{tab:classical} };
\item Ponderomotive squeezing or quantum back-action in the system.
\end{itemize}}
\textcolor{black}{Whether the joint contribution can be controlled at the level of $\bar{\dot{n}}_0 \Delta f \lesssim \sigma^2$ ultimately determines if the system can effectively saturate its quantum-limited performance.
}

\subsection{Channel Limit} \label{sec:channel}
In the previous two schemes, the input quantum state was fixed. However, even when allowing full optimization of the input quantum state, the ultimate precision for stochastic gravitational-wave detection remains fundamentally limited in the presence of losses. Specifically, if a signal is encoded through a quantum channel \( \Lambda_\sigma \), the Quantum Fisher Information (QFI) achievable by any quantum state is bounded by the Extended Channel QFI (ECQFI), which represents the maximum extractable information attainable using an ideal lossless ancilla channel and is defined as:
\begin{equation}
\mathcal{F}_Q^{\Lambda_\sigma}(\sigma)=\sup _{|\Psi\rangle} \mathcal{F}_Q^{\left(\Lambda_\sigma \otimes 11_A\right)(|\Psi\rangle\langle\Psi|)}(\sigma)\,,
\end{equation}
where \( 11_A \) denotes an ideal ancilla channel, and the supremum is taken over all possible input quantum states.

For a channel experiencing a pre-encoding loss \( \varepsilon \), the ECQFI for stochastic signals is given by\,\cite{Stochastic}:
\begin{equation}
\mathcal{F}_Q^{\Lambda_\sigma^{\text{noisy}}}(\sigma)=\frac{4}{2\varepsilon+\sigma^2}.
\end{equation}
Consequently, the fundamental limit of achievable SNR and the corresponding minimal detectable threshold become:
\begin{equation}
\begin{aligned}
{\rm SNR}_{\rm channel}&=\sqrt{\frac{N\sigma^2}{2\varepsilon}}=\frac{1}{\sqrt{\varepsilon}}{\rm SNR}_{\rm pc},\\
h^{2}_{\rm c,min}(f)&=\frac{2\varepsilon f}{ T_{\rm int}\Delta f \cdot \bar{S}^{\rm vac}_{\cal EE}(f)\overline{{a}^2}(f)}\,.
\end{aligned}
\end{equation}
This result, known as the \textit{channel limit}, sets the absolute bound for parameter estimation accuracy over all possible quantum states and measurement strategies {in detection system without internal squeezing}.  {\noindent
Meanwhile, Eq.~\ref{eq:sqz} indicates that the sensitivity of systems with internal squeezing is subject to a similar fundamental bound, with a simple substitution of $\varepsilon \to \varepsilon_b$. 
Thus, the channel limit in fact establishes an ultimate and unbreakable sensitivity bound for general detection systems.
}Regardless of the measurement employed, surpassing this precision is fundamentally impossible.

However, the exact conditions required to saturate this limit remain unclear. In the ideal (lossless ancilla) high-energy limit, two-mode squeezed vacuum (TMSV) states can achieve the ECQFI. In realistic lossy ancilla channels, certain non-Gaussian states may offer greater robustness compared to TMSV states, although states that precisely saturate the ECQFI in practical conditions are still under active research. It is clear, however, that entanglement with an ancilla system is necessary to approach this bound. Given that neither the optimal state preparation nor the readout scheme is currently known, reaching this channel limit is presently viewed as highly futuristic and challenging to achieve in the near term.

\section{Numerical Modeling of Sensitivity}

In this section, we explain the numerical sensitivity curves in Fig.\,1 of the main text, outlining each proposal’s concept, layout, and key parameters in numerical modeling.

\subsection{Linear Measurement}\label{sec:linear}
{We firstly study the performance of three individual proposals with linear measurement in Fig.\,3 in the main text which can be unified by the unified framework we developed.}

\noindent\textbf{Laser interferometer:} The most established linear gravitational wave detection system is the ground-based laser interferometer. Among various high-frequency designs of interferometer, we focus on the L-shaped resonator (also known as Fox-Smith configuration)\cite{L-shaped}, which maintains a high antenna response at the arm's first optical resonance by folding the arm cavity.

Optical gain of this detection system is given by:

\begin{equation}\label{homodyne_gain_2}
|G(f)|^2=2\sum_k\frac{2\gamma_{\rm vac} \omega_0 \bar{\cal E}\overline{{a}^2}(f)}{\hbar  \left[(2\pi f-\Delta_k)^2+\gamma_{\rm vac}^2\right]}\,.
\end{equation}
In the system, the energy is stored in the form of circulating laser power inside arm cavity, $\bar{\cal E}=\frac{2P_{\rm arc}L}{c}$ , where $L$ is the length of each fold of L-shaped cavity, $\omega_0$ is the angular frequency of laser. Meanwhile, $\Delta_k=\frac{\pi(2k-1)c}{2L}$  and $\gamma_{\rm vac}=\frac{cT_{src}}{2L}$ is the angular frequency and bandwidth of the k-th optical resonance, respectively. Here, $T_{\rm src}$ represents the effective transmissivity of the signal recycling cavity, formed by the initial test mass(ITM) and signal recycling mirror(SRM). Here, the extra factor of 2 is due to the resonant modes of this system distributes symmetrically around the carrier.

As a standard linear measurement, the detection sensitivity of this proposal is fundamentally loss-limited. Quantum loss inside the system mainly comes from arm loss (with $\Delta_k=\frac{\pi(2k-1)c}{2L}$, $\gamma_{k} = \frac{cT_{\rm src}}{2L}$, ${\cal L}_{\rm k}=\varepsilon_{\rm arm}/2T_{\rm itm}$ in Eq.\,\ref{Lossrate}) and SRC loss ($\Delta_k=\frac{\pi(2k-1)c}{2L}$, $\gamma_{k} = \frac{cT_{\rm src}}{2L}$, ${\cal L}_{\rm k}=\varepsilon_{\rm src}\cdot \frac{T_{\rm itm}}{4T_{\rm src}}$), where $T_{\rm itm}$ represents the transmissivity of ITM.
Near optical resonance, the loss rate can be be estimated as:\begin{equation}\label{loss} \varepsilon(f)\approx \frac{T_{\rm itm}}{T_{\rm src}}\varepsilon_{\rm src}+\frac{2}{T_{\rm src}}\varepsilon_{\rm arm},. \end{equation} At high frequencies, the arm loss dominates.

The sky-averaged antenna response $\overline{{a}^2}(f)$, computed similarly to LIGO-like detectors\cite{prd17}, peaks at cavity resonances $f=(k+\frac{1}{2})f_{0}$. At high frequencies, these peaks of antenna response asymptotically scales as: \begin{equation}\label{antenna_estimate} \overline{{a}^2}_{\rm max}(f) \approx 0.090\left(\frac{f_0}{f}\right)^2\,, \end{equation} where $f_0=c/2L$. The first resonance peak at $f=f_0/2$ has: \begin{equation}\label{antenna_first_1} \overline{{a}_1^2}\approx 0.156\,. \end{equation}

\begin{figure}[htbp] \centering \includegraphics[width=0.9\linewidth]{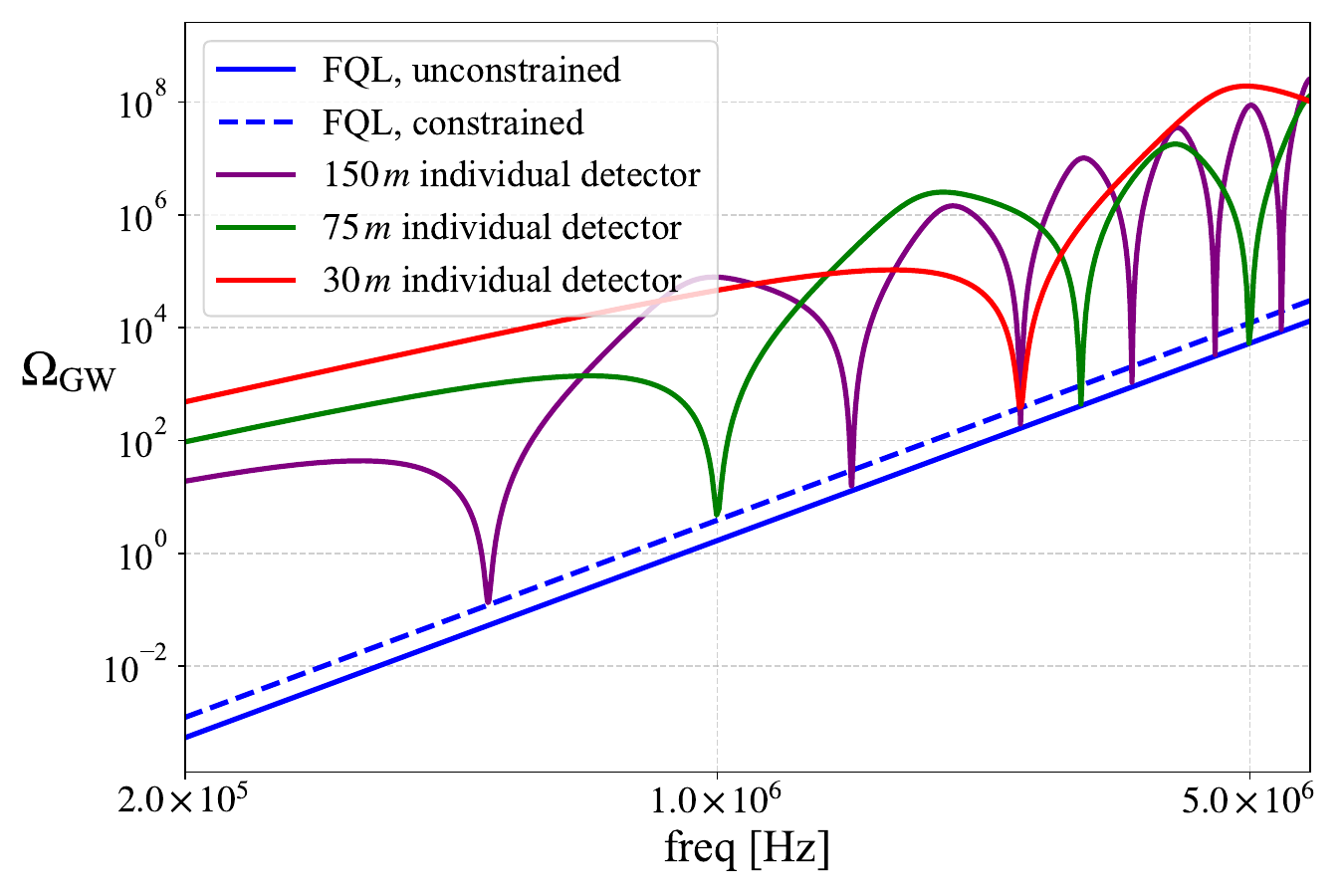} \caption{\textcolor{black}{Constrained and unconstrained  fundamental limit for linear readout within laser interferometer, with $T_{\rm int}=10\, \rm y$ and $\Delta f= 1\,\rm kHz$. Individual detectors with different lengths saturates the fundamental limit at different frequencies, similar to the standard quantum limit (where interferometers with varying circulating power touch the bound at different frequencies).}} \label{fig:Omega_GW} \end{figure}

The parameters we utilized to model the detectable threshold in Fig.\,1 in the main text are listed in Table.\,\ref{table:L-resonators}. In Fig.\,\ref{fig:Omega_GW}, we provide the performance of a single L-shaped interferometer. Clearly, single detector achieves optimal sensitivity only within a narrow bandwidth around resonance, to saturate the broadband detectable threshold in a wide frequency range, combining the peak sensitivity of multiple individual detectors is required. This applies to other detection proposals as well.

With different optical parameters, the broadband $\Omega_{\rm GW}$ detectable threshold of interferometer with linear readout is numerically given by:

\begin{small}
\begin{equation}\label{homodyne_limit}
\begin{aligned}
\Omega_{\mathrm{GW,min}}^{\rm laser}(f)&=1.7\times \left(\frac{10\mathrm{MW}}{P_{\mathrm{arm}}}\right) 
\left( \frac{\varepsilon_{\mathrm{arm}}+\frac{\varepsilon_{\mathrm{src}}T_{\mathrm{itm}}}{2}}{(10+\frac{100\times0.01}{2})ppm}\right)\\
&\times\left( \frac{10y}{T_{\mathrm{int}}} \right)^{\frac{1}{2}} 
\left( \frac{1\mathrm{kHz}}{\Delta f} \right)^{\frac{1}{2}}
 \left( \frac{f}{1\mathrm{MHz}}\right)^5  \,.
 \end{aligned}
\end{equation}
\end{small}
\textcolor{black}{The unconstrained bound corresponds to the numerical fundamental quantum limit in the main text.}

For detection restricted to the first resonance frequency, due to the different antenna response, the minimum detectable threshold would be modified as: \begin{small}
\begin{equation}\label{homodyne_limit_first}
\begin{aligned}
\left(\Omega^{\rm laser}_{\mathrm{GW,min}}\right)^1(f)&=3.9\times \left(\frac{10\mathrm{MW}}{P_{\mathrm{arm}}}\right)
\left( \frac{\varepsilon_{\mathrm{arm}}+\frac{\varepsilon_{\mathrm{src}}T_{\mathrm{itm}}}{2}}{(10+\frac{100\times0.01}{2})ppm}\right)
\\
&\times
\left( \frac{10y}{T_{\mathrm{int}}} \right)^{\frac{1}{2}} 
\left( \frac{1\mathrm{kHz}}{\Delta f} \right)^{\frac{1}{2}}
 \left( \frac{f}{1\mathrm{MHz}}\right)^5  \,.
 \end{aligned}
\end{equation}
\end{small}
\textcolor{black}{FQL benchmarks in Fig.\,2 in the main text correspond to the constrained version of fundamental limit (with $\Delta f=\gamma$ ).}
\begin{table*}
\begin{center}
\begin{threeparttable}
\caption{Key parameters in the numerical calculation of $\Omega_{\mathrm{GW}}$ detecting limit.  Here, parameters here is largely better than the ultimate targets of third-generation gravitational wave detectors, thus it only gives a futuristic limitation for the fundamental sensitivity level. }
\par
\begin{tabular}{cccc}
\hline
Symbol& Description& Value &Unit\\
\hline
$\omega_0$&\quad Laser frequency &  $1.77\times 10^{15} $ & Hz\\
$\lambda_0$&\quad Laser wavelength & $1064$ &nm \\
$P_{\mathrm{arm}}$&\quad Circulating power inside arm cavity& 10 & MW\\
\hline
$T_{\mathrm{itm}}$&\quad Transmissivity of ITM & 0.01& \\
$T_{\mathrm{src}}$&\quad Transmissivity of signal recycling cavity& 0.01& \\
&\quad(Almost entirely transmissive signal recycling mirror)& & \\
\hline
$\varepsilon_{\mathrm{arm}}$ &\quad Arm loss coefficient & 10 & ppm \\
$\varepsilon_{\mathrm{src}}$ &\quad Src loss coefficient &\hspace{1cm} 100 \hspace{1cm} & ppm \\
\hline
\label{table:L-resonators}
\end{tabular}
\end{threeparttable}
\end{center}
\end{table*}

\noindent\textbf{EM cavity:} Detecting high-frequency gravitational wave signals using a resonant electromagnetic cavity is a newly proposed scheme\cite{EMcav}. In this proposal, the energy is stored a strong static magnetic field, and the gravitational wave signal alters the total magnetic energy stored within the cavity when it passes by. However, previous noise estimations for this proposal only accounted for readout noise, leading to an overestimation of its performance. Here, we provide a more reliable estimation of the fundamental quantum limit for this design.

{This system also closely follows the energetic quantum limit, with the carrier being static magnetic field. Optical gain of this proposal is given by:}
\begin{equation}
|G(f)|^2 = \frac{\Delta \gamma B_0^2V\overline{{a}^2}(f)}{\mu_0\hbar \cdot\left[(2\pi f-\Delta)^2+\gamma^2\right] } \,,
\end{equation}
{where $B_0$ is the strength of static magnetic field, $\Delta$ and $\gamma$ is the angular frequency and bandwidth of the first cavity resonant mode.} 


Due to the outgoing field is centered at the radio and microwave frequency bands, squeezing and photon counting are challenging to implement in this proposal. Consequently, in the modeling, we select vacuum homodyne as the readout method, { with the static magnetic field itself being the carrier, and the direct measurement quantity is the total power within the cavity.} The SNR level and minimum detectable threshold are given by:

\begin{equation}\label{QCRB_vachom}   
{\rm SNR}\left[ \sigma^2 \right]={\sigma^2} \sqrt{T_{\rm int}\Delta f}\,,
\end{equation}
\begin{equation} 
\begin{aligned}\label{threshold_vachom}
h^{2}_{\rm c,min}(f)&=\frac{f}{ \sqrt{T_{int}\Delta f} \cdot |G(f)|^2}\\
\Omega_{\rm GW,min} &=\frac{4\pi^2}{3H_0^2}\frac{f^3}{ \sqrt{T_{int}\Delta f} \cdot |G(f)|^2}\,.
\end{aligned}
\end{equation}

In the modeling of detectable threshold, we  set the static magnetic field as 5T, and estimate detection bandwidth \( \gamma \)  as a frequency independent constant of $1\, \rm kHz$. For the sky-averaged antenna response, which is not yet well-studied, we estimated it as $\overline{{a}^2}(f)\sim 1$, which would not introduce order-of-magnitude difference to the fundamental limit. In Fig.\,\ref{fig:limits} in the main text, we present two specific benchmarks to numerically illustrate the sensitivity limits of EM cavity designs:

\begin{itemize}
\item {\textbf{Fixed aspect ratio case:} In this benchmark, we consider a cylindrical cavity with aspect ratio of $1:5$, identical to the settings in Ref.\,\cite{EMcav}. Under this aspect ratio assumption, the size of detector should closely follow the central detection frequency, \( V \propto L^3 \propto f^{-3} \) . The numerical minimum detectable threshold for the fixed-aspect-ratio case and its scaling behavior with classical parameters are given by:}
\begin{equation}\label{cav_limit}
\begin{aligned}
\Omega_{\mathrm{GW,min}}^{\rm cav}(f)&=6.7\times 10^{-4}\times \left(\frac{5\mathrm{T}}{B_{{0}}}\right)^2
\left( \frac{\gamma}{1\, \rm kHz} \right)\\
&\times\left( \frac{10y}{T_{\mathrm{int}}} \right)^{1/2}
\left( \frac{1\mathrm{kHz}}{\Delta f} \right)^{1/2}
 \left( \frac{f}{1\mathrm{MHz}}\right)^5  \,.
 \end{aligned}
\end{equation}
Even with the absence of squeezing, the quantum limit of EM cavity still largely exceeding the loss limit of interferometer. That is due to the immense energy stored in static magnetic field actually brings a far superior energetic quantum limit. However, large energy storage also put additional challenge on the feasibility of this proposal, especially at kHz range.
For instance, setting the central frequency to 100 kHz would require a {kilometer-scale}, along with immense energy storage which is almost impossible to achieve. 
The MHz and GHz frequency ranges are more likely to be within the scope of this {fixed aspect ratio scenario}.
\item {\textbf{Fixed volume case:} A more practically useful benchmark is the performance of an EM cavity with a fixed total volume, since a fixed volume corresponds to manageable construction and operational costs. 
Here, we consider a rectangular-cuboid resonant cavity. By varying its aspect ratio, the frequency of its fundamental resonant mode can, in principle, be shifted to arbitrarily high values.}

{Assuming that the antenna response remains of order unity across different aspect ratios, the numerical minimum detectable threshold and scaling behavior for the fixed‑volume case is given by: }
{\begin{equation}\label{cav_limit2}
\begin{aligned}
\Omega_{\mathrm{GW,min}}^{\rm cav}(f)&=2.7\times10^{-5}\times \left(\frac{5\mathrm{T}}{B_{{0}}}\right)^2
\left( \frac{\gamma}{\rm 1kHz} \right)\\
\times\left( \frac{10y}{T_{\mathrm{int}}} \right)^{1/2}
&\left( \frac{1\mathrm{kHz}}{\Delta f} \right)^{1/2}
\left( \frac{[300 \, \rm m]^3}{V} \right)
 \left( \frac{f}{1\mathrm{MHz}}\right)^2  \,.
 \end{aligned}
\end{equation}}

{Noticing that, for a fixed volume $V = L^3$, this fixed-volume bound applies only in the frequency range
$f \ge \sqrt{3}\,c/L$, which corresponds to the range over which the resonant frequency of the fundamental mode can be accessed by varying the cavity aspect ratio.}
\end{itemize}

{Meanwhile, for EM cavity systems operating in the radio and microwave bands, the dominant classical noise is thermal noise. The system can reach the quantum limit only if the thermal background is suppressed below the vacuum quantum fluctuations, $k_B T\lesssim 2\pi \hbar f_0$, where $T$ and $f_0$ denotes the temperature within cavity and targeting frequency. This imposes a stringent requirement on the cooling within the cavity:
\begin{equation}
T\lesssim 48 {\rm \mu K} \times\frac{f_0}{1\, \rm MHz}\,.
\end{equation}With current cooling technology at the $\mathrm{mK}$ level, the EM cavity design would be thermally limited in $\rm kHz$ and $\rm MHz$ bands; the quantum limit can only be achieved with large‑scale cooling at $\mu\mathrm{K}$ or lower temperatures, which requires further advancements in cooling and cryogenic technology.}

\vspace{5pt}

\noindent\textbf{Levitating sphere:} Levitating sphere is an  {intriguing approach} for high-frequency detection\cite{Levsphere_1,Levsphere_2}.  {The basic idea is to levitate a dielectric material inside the arm cavity via optical tweezers, thus manipulating dynamical back action and enhancing sensitivity to high-frequency gravitational wave signals. \textcolor{black}{For a fixed levitated object inside the cavity, the laser power required to achieve infinite ponderomotive squeezing at each frequency is uniquely determined, leaving the mass of the object $m_{\rm lev}$ as the primary free parameter.
}

Existing estimation of performance for this proposal primarily focus on brownian thermal noise and mechanical dissipation of the dielectric object, and haven't considered the quantum noise of electromagnetic field inside the system. Here, we provide a simple estimation of the fundamental quantum limit for the levitating sphere.

{Unlike other proposals, the optical tweezer effect creates strong mechanical back action in the system. However, as illustrated in the quantum loss section, as long as the scheme is loss-limited, quantum loss that are introduced before signal encoding cannot be recovered. In this proposal, part of the arm loss is introduced before mechanical back action, which would become the ultimate limit of the system. Expression of arm loss is identical to Eq.\,\ref{loss}.

In the modeling, we set the loss rate of initial mirror and the dielectric object as both $10 ppm$. For the mass of dielectric material and length or finesse of cavity, we adopt the level in  \cite{Levsphere_2}.
Meanwhile, due to the optical tweezer effect, the arm circulating power is determined the mass of levitating object, thus cannot reach the level of laser interferometer. Under the parameter settings considered here, the circulating power is limited at $0.1$ W to $10$ kW in the frequency range we considered. Meanwhile, for order-of-magnitude estimation, the antenna response in this case is roughly estimated as $\overline{{a}^2}=1$.

As shown in Fig.1 in the main text, from the perspective of quantum limit, levitating sphere shows no clear advantage over traditional designs, primarily restricted by the arm circulating power. The main advantage of this design is its ability to achieve high level of squeezing. In current laser interferometers, the quantum noise is primarily restricted by the squeezing level we could achieve, a broad-band, 10dB squeezing is still not achievable yet. The strong internal squeezing in levitating sphere provides a valid solution to this technical restriction, which makes it also an alternative for future detectors.

\subsection{Different Measurement Scheme}

The rest two proposals, photon counting and channel limit, basically have physical layout of L-shaped laser interferometer , but differs in the measurement scheme. 

\vspace{5pt}

\noindent\textbf{Photon Counting:}
Basic principle and concept of photon counting is discussed in the metrology section. Here, we numerically study its performance in ideal condition with $\bar{\dot{n}}=0$,  corresponding to the case without classical noise, dark count, and mechanical back action. In the modeling in the main text, we identically utilized the optical parameters in Table.\,\ref{table:L-resonators}. 

{Broadband detectable \(\Omega_{\mathrm{GW}}\) threshold for ideal photon counting could be numerically estimated as:}
\begin{equation}\label{photoncounting_limit_ideal}
\begin{aligned}
\Omega_{\mathrm{GW,min}}^{\rm pc}(f)&=2.8(6.6)\times 10^{-3} \times\left(\frac{10\mathrm{MW}}{P_{\mathrm{arm}}}\right) 
\left( \frac{T_{\mathrm{src}}}{0.01} \right)\\
&\times\left( \frac{10y}{T_{\mathrm{int}}} \right)
\left( \frac{1\mathrm{kHz}}{\Delta f} \right)
 \left( \frac{f}{1\mathrm{MHz}}\right)^5.
\end{aligned}
\end{equation}}
{Here, the value within parentheses correspond to case when only the first optical bandwidth is employed.}

\vspace{0.3cm}
\noindent\textbf{Channel Limit:}
Channel limit is actually the ultimate sensitivity limit over all the possible input quantum state and readout scheme, and specific measurement scheme that saturates the channel limit remains largely uncertain. Numerical value for minimum detectable threshold for \( h_c^2 \) and $\Omega_{\rm GW}$ for channel limit are simply \( {\varepsilon} \) times of that of ideal photon counting, with the same loss rate as interferometer with linear measurement. 

Due to the use of entangled and non-Gaussian states, saturation of channel limit would be even more challenging and futuristic compared to linear readout and photon counting. The loss-limited behavior highlights the extensive manipulation required for the input quantum state, posing challenges for state preparation. Similarly, the ideal performance with zero noise in the absence of a signal implies that the corresponding quantum state and measurement would be highly sensitive to ancilla loss and internal squeezing.
Nonetheless, as the ultimate limit across all possible metrology, it is still important in guiding and constraining future designs.

{\section{Practical Design Trade}}
In this section, we discuss how the fundamental limit guides the design of future detectors. 
In general, any detection scheme consists of two essential ingredients: the physical probe system and the associated metrology. 
Preliminary design of a complete detection scheme can be hierarchically divided into four stages, with the fundamental limit providing guidance at each step. 
Below, we present these stages alongside their application in the worked example of detecting the domain-wall gravitational-wave signal (illustrated in Fig.\,1 of the main text), aiming for the highest possible SNR around $75\,\rm kHz$.

\begin{itemize}
\item[(a)] \textbf{Fixing the physical system:}  
The first step is to identify the most suitable physical system among the many candidate platforms. At this stage, the role of the fundamental limit is primarily heuristic: it does not prescribe how to build a device, but rather highlights a key design insight—namely, that a promising probe system must support either large energy storage, a high-$Q$ cavity resonance mode, or preferably both.

For the worked example, candidate systems to probe the benchmark signal include laser interferometers and electromagnetic (EM) cavities. Both exhibit quantum potential for ultra-high-frequency GW detection: laser interferometers have well-calibrated resonant modes with high $Q$, while EM cavities offer substantial energy storage. Practical feasibility further helps narrow down the choice of physical system. On one hand, the typical detector size required to probe a $75\,\rm kHz$ signal is kilometer-scale; such a size for a laser interferometer is technologically mature, whereas constructing a three-dimensional resonant cavity at this scale would involve enormous cost and immature technology. On the other hand, laser interferometers benefit from well-established visible and near-infrared detection techniques and are compatible with a variety of advanced metrology, whereas resonant cavities operating at radio-wave frequencies are limited to linear measurements and cannot implement advanced techniques such as photon counting. Based on these considerations, a laser interferometer tends to be the optimal choice of physical system.

\item[(b)] \textbf{Selecting suitable metrology:}  
Once a physical system is chosen, the most suitable metrology is examined. Since advanced metrology often involves greater technical complexity, the simplest approach that satisfies the sensitivity requirements is typically chosen for practical implementation.

Within the working example, options for the laser interferometer include linear readout, photon counting, or entanglement-based advanced schemes, all fundamentally bounded by the channel limit. Comparing their performance with the domain-wall benchmark signal (Fig.\,1) shows that photon counting provides sufficient quantum reach with state-of-the-art parameters. Linear readout requires unrealistically low losses, and advanced metrology that saturate the channel limit introduce additional implementation complexity, while the extra quantum potential they provide is not essential for this particular case. Therefore, photon counting is selected as the practical choice.

\item[(c)] \textbf{Optimizing the fundamental limit:}  
With the system and metrology fixed, the fundamental limit can be further optimized by tuning key parameters: total energy $\bar{\mathcal{E}}$, resonant bandwidth $\gamma$, antenna response $a$, and loss coefficient $\varepsilon$. For photon counting within laser interferometer, the optimization procedure is as follows:
\begin{itemize}
    \item \textit{Energy storage $\bar{\mathcal{E}}$}: The SNR scales proportionally with $\bar{\cal E}$, so larger energy or higher circulating power improves sensitivity. We select $\bar{\cal E}=10\,\rm MW$, a near-future practical level.
    \item \textit{Resonant bandwidth $\gamma$(or equivalently, $Q$)}: The SNR is insensitive to $\gamma$ for broadband signals, due to the gain-bandwidth trade-off: increasing $\gamma$ widens the effective bandwidth but reduces peak sensitivity proportionally, leaving the ultimate SNR of single detector unchanged.
    \item \textit{Antenna response $a$}: The SNR is proportional to $a$. Optimizing $a$ involves adjustment of geometric configuration; the L-shaped resonator design achieves optimal high-frequency response\,\cite{L-shaped}.
    \item \textit{Loss coefficient $\varepsilon$}: Photon counting is insensitive to quantum loss, as the loss channel does not generate additional photons.
\end{itemize}
In this case, the optimization procedure guides us to select the L-shaped interferometer as the configuration for detecting signals around $75\,\rm kHz$. For other scientific targets, the same procedure can be applied to determine the appropriate configuration or working point.

\item[(d)] \textbf{Constraining classical noise:}  
The gap between the fundamental quantum limit and the required sensitivity can be translated into qualitative constraints on classical noise. For photon counting, such classical noise can be quantified by the residual photon flux $\bar{\dot{n}}_0$, which further degrades the achievable sensitivity. Fig.\,\ref{fig:time} shows the required integration time to reach unity SNR for the target signal under various $\bar{\dot{n}}_0$ levels. The current dark-count rate ($\bar{\dot{n}}_0 \approx 6.6 \times 10^{-6}\,\rm s^{-1}$) exceeds the requirement ($\bar{\dot{n}}_0 \le 0.8 \times 10^{-6}\,\rm s^{-1}$), indicating that further reduction is needed for practical implementation.
\end{itemize}

\begin{figure}[htbp] 
\centering 
\includegraphics[width=0.9\linewidth]{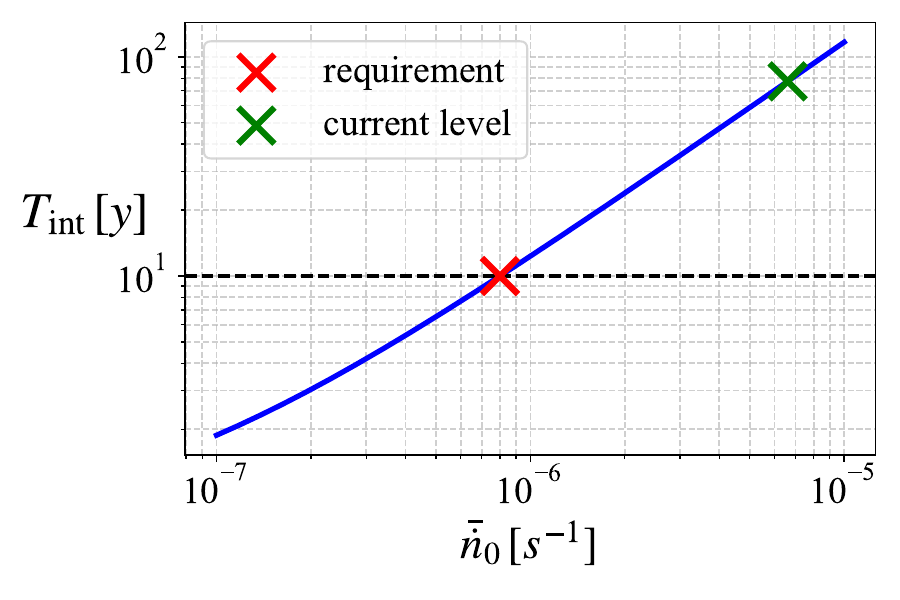} 
\caption{Minimum integration time to achieve unity SNR under different $\bar{\dot{n}}_0$ levels.}
\label{fig:time} 
\end{figure}

In summary, the fundamental quantum limit provides a universal bound that guides detector design systematically. By translating scientific goals into design steps, the framework allows practical optimization of sensitivity and constraining classical noise, thus establishing a clear path for future ultra-high-frequency gravitational-wave detectors.

\subsection{{Sources of ultrahigh-frequency GWs in SO(10) GUTs}}

{The GUT framework naturally predicts new physics occurring at scales far above the electroweak scale, offering rich sources for generating ultra-high frequency GWs. While the minimal GUT, Georgi-Glashow (GG) $SU(5)$ \cite{Georgi:1974sy}, does not involve any intermediate symmetry breaking, we will focus on the next-to-minimal GUT framework, namely $SO(10)$, which provides rich phenomenologies in particle physics and cosmology \cite{Fritzsch:1974nn}. Since $SO(10)$ GUTs already encompass most of the interesting features presented in larger GUT groups, we will not extend our analysis beyond this point. With a systematic analysis of all symmetry breaking chains within $SO(10)$ GUTs, we demonstrate that high frequency GWs are naturally predicted in many GUT models. 
Our focus will be on two primary GW sources: GUT-motivated cosmological phase transition and metastable cosmic strings.  
Although stable gauge strings are also widely predicted in GUTs, they will not be considered here, as their gravitational wave spectra do not exhibit distinctive features in the high-frequency regime compared to those in the Hz band \cite{King:2020hyd}.} 

There are various ways to achieve the spontaneous breaking from $SO(10)$ to the SM gauge group. 
Each ``breaking chain'' has a distinct pattern of intermediate gauge symmetries and we use the following abbreviations for these
gauge groups:
\begin{align} \label{eq:symmetries}
G_{51} &= SU(5) \times U(1)_\chi \,, \nonumber\\
G_{51}^{\rm flip} &= SU(5)^{\rm flip} \times U(1)_V \,, \nonumber\\
G_{422} &= SU(4)_c \times SU(2)_L \times SU(2)_R \,,\nonumber\\
G_{421} &= SU(4)_c \times SU(2)_L \times U(1)_R\,, \nonumber\\
G_{3221} &= SU(3)_c \times SU(2)_L \times SU(2)_R \times U(1)_X \,, \nonumber\\
G_{3211} &= SU(3)_c \times SU(2)_L \times U(1)_Y \times U(1)_{B-L} \,, \nonumber\\
G_{\rm SM} &= SU(3)_c \times SU(2)_L \times U(1)_Y\,. 
\end{align}
Here, $SU(5)$ represents the usual GG group, $SU(5)^{\rm flip} \times U(1)_V$ is the gauge group in the flipped $SU(5)$ model, in which the up-type quarks and down-type quarks, as well as neutrinos and charged leptons, are flipped in their representations compared with those in the GG $SU(5)$. $G_{422}$ is the Pati-Salam (PS) gauge group, where leptons are treated as the fourth colour of quarks. $SO(10)$ includes an intrinsic $Z_2^C$ parity symmetry between the left chiral field and the charge conjugate of the right chiral field. It might remain unbroken during $SO(10)$ breaking to $G_{422}$. In the PS model, this $Z_2^C$ appears as the permutation symmetry between matter fields $(\overline{\mathbf{4}}, \mathbf{2}, \mathbf{1})$ and $(\mathbf{4},\mathbf{1},\mathbf{2})$.   
We abbreviate the PS gauge group preserving $Z_2^C$ as $G_{422}^C \equiv G_{422} \times Z_2^C$. 
This $Z_2^C$ may be still preserved for the breaking from $G_{422}$ to $G_{3221}$, and we denote the latter case as $G_{3221}^C \equiv G_{3221} \times Z_2^C$. 
Note that in some notation, the $G_{3211}$ is considered to be $SU(3)_c \times SU(2)_L \times U(1)_R \times U(1)_X$. It is equivalent to represent as $ U(1)_{Y} \times U(1)_{B-L}$ as the former two $U(1)$ charges are just a mixing of the later two, i.e., $\pm R = Y - \frac{B-L}{2}$ and $X = \sqrt{\frac{3}{2}}\frac{B-L}{2}$, where the $\pm$ signs refer to two ways of embedding $G_{3211}$ in $G_{51}$ and $G_{51}^{\rm flip}$. 


All breaking chains from $SO(10)$ to $G_{\rm SM}$ are presented in Fig.~2 in the main text, which can be classified into four categories \cite{King:2020hyd} and also presented below. Each breaking chain leads to topological defects generated in the early Universe. {We systematically assess, for each breaking chain, the potential observability of a stochastic gravitational wave background (SGWB) signal generated by a cosmological phase transition or metastable strings, with particular focus on the phase transition resulting from intermediate symmetry breaking in each chain.}  
\begin{itemize}
\item[(a)] The breaking chain with the GG $SU(5)$ and the $U(1)_{B-L}$ as an intermediate symmetry.
\begin{align}
    SO(10) \chain{\rm m}{} G_{51} \chain{\rm m}{} G_{3211} \chain{\rm s}{} G_{\rm SM} \nonumber
\end{align}
The breaking of $SO(10) \to G_{51}$ and $G_{51} \to G_{3211}$ (which is essentially $SU(5) \to G_{\rm SM}$) generates monopoles, which are denoted by ``m'' above the arrow. The null observation of proton decay requires the baryon number violating (BNV) breaking scale, along with the corresponding monopole mass generated at the same scale, to be above $10^{15}$~GeV. These super heavy monopoles will dominate the energy density of the Universe later during the expansion in the radiation era and thus they are unwanted topological defects and should be inflated. The breaking $G_{3211} \to G_{\rm SM}$ (which is essentially the breaking of the gauge $U(1)_{B-L}$) generates cosmic strings, denoted as ``s'' above the arrow. As the existence of cosmic strings is safe for the evolution of Universe in the radiation- and matter-domination eras, the inflation can be introduced before the breaking of $G_{3211}$. In other word, we can assume reheating processes before the breaking of $G_{3211}$. As a consequence, the phase transition for $U_{B-L}$ breaking and the generation of cosmic strings happens in the radiation era. In this breaking chain, there is no restriction on the breaking scale of $U(1)_{B-L}$ as it is separated from the $SU(5)$ unification. For a SGWB spectrum with the peak frequency around kHz - MHz, the $U(1)_{B-L}$ breaking scale should be somewhere within $10^6$ - $10^9$~GeV. 

\item[(b)] The breaking chain with the flipped $SU(5)\times U(1)$ as intermediate symmetry. 
\begin{align}
    SO(10) \chain{\rm m}{} G_{51}^{\rm flip} \chain{\rm s}{} G_{\rm SM} \nonumber
\end{align}
In this chain, monopoles from the first step breaking should be inflated away, and the breaking of $G_{51}^{\rm flip} \to G_{\rm SM}$ produces cosmic strings which are safe for the cosmological evolution. All breaking scales should be $\gtrsim 10^{15}$~GeV to satisfy the proton decay constraint. If a SGWB is generated via phase transition, the frequency will be super-high with the peak frequency roughly around $10^{15}$~Hz, which is impossible to be observed in the future GW measurements. {On the other hand, since the two breaking scales are naturally nearby, a small $\kappa$ is predicted and observable GWs via metastable strings are likely produced in this chain.}

\item[(c)] Breaking chains with the PS symmetry $G_{422}$ or its subgroups as intermediate symmetries. This category provides the largest possibility with 31 different chains in total \cite{King:2021gmj}.
They can be further sorted within two types.

\begin{align}
    (\text{c}_1) &\quad 
    SO(10) \to ...\to G_{421}, G_{3221}, G_{3211} \chain{\rm s}{} G_{\rm SM} \nonumber \\
    (\text{c}_2) &\quad 
    SO(10)  \to ... \to G_{422} \chain{\rm m}{} G_{\rm SM} \nonumber \\
    (\text{c}_3) &\quad  
    SO(10)  \to ... \to G_{422}^{C}, G_{3221}^{C} \chain{\rm w, (s,m)}{} G_{\rm SM} 
\end{align}

In (c$_1$), the spontaneous breaking of $G_{421}$, $G_{3221}$ and $G_{3211}$, along with the subsequent breaking of their subgroups, leads to the formation of topological defects no more than cosmic strings. As a result, inflation can be introduced before it. 
{We further categorize the following chains as ($\text{c}_{11}$),
\begin{align}
& ...\to  G_{422} \chain{\rm m}{} G_{421}, G_{3221}, G_{3211}  \chain{\rm s}{} G_{\rm SM}
\nonumber\\
& ...\to G_{421}, G_{3221} \chain{\rm m}{} G_{3211} \chain{\rm s}{} G_{\rm SM} \,.
\end{align}
In these chains, cosmic strings are formed during the final symmetry breaking step to the SM, while monopoles are generated during the preceding intermediate breaking step.} {In (c$_2$), the breaking of $G_{422}$, generates only monopoles.}  In (c$_3$), the breaking of the last intermediate symmetry to $G_{\rm SM}$ generates domain walls, denoted as ``w'' above the arrow. 
Most breaking chains (26 of 31) belong to type (c$_1$) and only 2 chain and 3 chain belongs types (c$_2$) and (c$_3$), respectively \cite{King:2021gmj}. 
{We analyze potential SGWB via phase transitions in these chains.} For type (c$_1$) chains, similar to chains in category (a), inflation can be introduced before the breaking of the lowest intermediate symmetry, and thus the SGWB from the phase transition, if it is first-order, might be observed. Note that, here the phase transition could happen along with $G_{3221} \to G_{\rm SM}$ or $G_{421} \to G_{\rm SM}$, which are essentially the breaking of $SU(2)_R \times U(1)_X \to U(1)_Y$ and $SU(4)_c \times U(1)_R \to SU(3)_c \times U(1)_Y$, respectively. More DOF of gauge bosons, as well as necessary Higgs multiplets responsible for the symmetry breaking, helps to promote the phase transition to be first-order. Following the discussions in e.g., \cite{Bertolini:2009qj,Chakrabortty:2019fov,King:2021gmj}, the energy scale of the lowest intermediate symmetry breaking span several orders of magnitude. It could be as high as $10^{13}$~GeV, or decrease to $10^{6}$~GeV, depending on the breaking chains and particle contents included in the RG running for the gauge unification. In this work, we will take care those might be observable in kHz - MHz GW measurements, referring to the energy scale around $10^6$ - $10^9$~GeV. %
In type (c$_2$), monopoles are generated, and their masses are naturally around the same scale of the symmetry breaking, i.e., in our preset regime $\sim 10^6$ - $10^9$~GeV. They are too light to dominate the energy density of the expanding Universe, and thus inflation does not have to be introduced after their production \cite{Lazarides:1984pq, Kephart:2001ix}. %
In type (c$_3$), domain walls are generated due to the spontaneous breaking of $Z_2^C$. Any topologically stable domain walls with the mass scale above 10~MeV generated after inflation can overclose the Universe \cite{Vilenkin:1984ib} and thus type (c$_3$) will not be considered. {As for metastable strings, we expect chains in category ($\text{c}_{11}$) predict such topological defects if the last two breaking scales are sufficiently close. It is worth noting that, in contrast to phase-transition-induced gravitational waves, an observable GW signal from metastable strings requires these breaking scales to lie not far below the GUT scale. Other chains, including the remaining ones in category ($\text{c}_{1}$) and those in categories ($\text{c}_{2}$) and ($\text{c}_{3}$), are not expected to produce observable metastable strings for the following reasons: the remaining ($\text{c}_{1}$) chains, namely, those including $Z_2^C$ in the second last step breaking, generate domain walls, which are severely constrained in cosmology; chains in category ($\text{c}_{2}$) have no strings generated at the final breaking step; 
and chains in ($\text{c}_{3}$) lead to domain walls formed during the last step of symmetry breaking.}

\item[(d)] Breaking with the standard $SU(5)$ subgroup as the lowest intermediate scale before breaking to $G_{\rm SM}$, 
\begin{align}
    SO(10) \to ... \to SU(5) \chain{\rm m}{} G_{\rm SM} \nonumber
\end{align}
These chains have the last intermediate symmetry as the GG $SU(5)$. The null observation of proton decay requires the $SU(5)$ scale above $10^{15}$~GeV. It is well-known that monopole with mass at such high scale has cosmological problems. Thus, inflation is necessary after the symmetry breaking. 

\end{itemize}
We find that most symmetry breaking chains in the $SO(10)$ framework---specifically types (a), (c$_1$), and (c$_2$)---allowing for cosmological phase transitions occurring well above the electroweak scale. These transitions avoid cosmological issues that would otherwise require inflation to resolve. Gravitational waves sourced by phase transitions in the radiation-dominated era may be detectable for any of these breaking chains. 
%
{In particular for chains (c$_{11}$), if the scale of the second lowest intermediate symmetry breaking and the associated monopole mass lie within the viable range discussed above, they do not lead to a cosmological problem. In this case, the corresponding phase transitions can be considered as potential sources of observable gravitational waves.
On the other hand, chains ($\text{b}$) and ($\text{c}_{11}$) can generate metastable strings, provided that the last two symmetry breaking scales are sufficiently close. Observable GW signals via such cosmic metastable strings are expected if these scales are sufficiently high.} 

In conclusion, the $SO(10)$ GUT framework provides rich possibilities for realizing cosmological phase transitions at scales in the range of $10^6$ - $10^9$~GeV during the radiation-dominated era. Depending on the specific symmetry breaking chains and the large number of degrees of freedom associated with the new particle contents involved in the phase transition, a first-order phase transition can occur. {In particular, certain chains with breaking scales slightly below the GUT scale may give rise to metastable cosmic strings. These well-motivated scenarios have the potential to generate sizable gravitational wave signals.}


\end{document}